\documentclass[11pt]{article}
\usepackage{latexsym}

\def\be{\begin{equation}}
\def\ee{\end{equation}}
\def\bea{\begin{eqnarray}}
\def\eea{\end{eqnarray}}
\def\pa{\partial}

\def\fn{\footnote}

\def\case#1/#2{\textstyle\frac{#1}{#2}}

\begin{document}
\begin{titlepage}
\vspace{.7in}

\begin{center}
\Large \bf{PDE System Approach to Large Extra Dimensions}
\\
\vspace{.7in} \normalsize \large{Edward Anderson and Reza Tavakol  }
\\
\normalfont
\normalsize
\vspace{.2in}
{\em  Astronomy Unit, School of Mathematical Sciences, \\
Queen Mary, Mile End Road, London E1 4NS, U.K. }

\begin{abstract}

We explore some foundational issues regarding the splitting of $D$-dimensional Einstein Field Equations 
(EFE's) with respect to timelike and spacelike ($D$ - 1)-dimensional 
hypersurfaces, first without and then with thin matter sheets such as branes.    
We begin to implement methodology, that is well-established for the GR Cauchy and initial value 
problems (CP and IVP), in the new field of GR-based braneworlds, identifying and comparing 
many different choices of procedure.  

We abridge fragmentary parts of the literature of embeddings, putting the Campbell--Magaard theorem into context.  
We recollect and refine arguments why York and not elimination methods are used for the GR IVP.    
We compile a list of numerous mathematical and physical impasses to using timelike splits, whereas spacelike 
splits are known to be well-behaved.  
We however pursue both options to make contact with the current braneworld literature which is almost entirely based on timelike splits.

In our study of timelike splits, we look at the Shiromizu--Maeda--Sasaki braneworld by means 
of reformulations which emphasize different aspects from the original formulation.  
We show that what remains of the York method in the timelike case generalizes heuristic bulk construction schemes.  
We formulate timelike (brane) versions of the thin sandwich conjecture.  
We discuss whether it is plausible to remove singularities by timelike embeddings.  
We point out how the braneworld geodesic postulates lead to futher difficulties with the notion of singularities 
than in GR where these postulates are simpler.

Having argued for the use of the spacelike split, we study how to progress to the construction 
of more general data sets for spaces partially bounded by branes.  Boundary conditions are 
found and algorithms provided.  Working with (finitely) thick branes would appear to 
facilitate such a study.  

\end{abstract}

\vspace{.2in}
\end{center}

\vspace{.3in}
Electronic address: eda@maths.qmw.ac.uk, reza@maths.qmul.ac.uk

\end{titlepage}

\section{Introduction}

A great deal of effort has recently gone into the study of 
higher-dimensional models as unified theories of gravity and 
fundamental matter fields.  The original 
interest in such models in Kaluza--Klein (KK) theory \cite{KK} was 
revived by the important role of dimension 11 
in supergravity \cite{sugy} and more recently by superstring theory
\cite{Polchinski} and M-theory \cite{Mtheory}, which favour spacetimes 
of dimension 10 and 11 respectively.  
The usual (KK) argument for the apparent 4-dimensionality of spacetime 
is that the extra dimensions are compactified i.e curled up sufficiently small so as not to 
conflict with observation.  However, recently much attention 
has been placed instead on string-theory-inspired models with 
large extra dimensions \cite{std, stdw1, stdw2, SMS,string1, string2, Turok}  
in which most of the physics is 
confined or closely-bound to a lower-dimensional braneworld
surrounded by a higher-dimensional bulk.  

Though motivated by superstring and M-theories, many of these models
\cite{std, stdw1, stdw2, SMS} are in fact formulated 
within the framework of (higher-dimensional) general 
relativity (GR).  Among these are:
1) the second Randall--Sundrum scenario \cite{stdw2} in which 
the graviton is tightly-bound to the brane by the curvature due to the 
warping of the bulk metric, which is pure anti-deSitter (AdS).  
2) A more general scheme of Shiromizu, Maeda and Sasaki (SMS) 
\cite{SMS}, in which the 4-dimensional Einstein field equations 
(EFE's) are replaced by 4-d ``braneworld EFE's''  
(BEFE's), which are not closed since there is a `dark energy' 
Weyl tensor term, knowledge of which requires solving also for the bulk. The interpretational 
difficulties due to this constitute the `Weyl problem'. 
The other main distinctive feature of SMS's BEFE's is the 
presence of a term quadratic in the braneworld energy-momentum, 
which arises from the junction condition used \cite{jns, SMS}.  

Two important questions arise in consideration of such models.  
{\bf Q1:} how should such models be built and interpreted consistently within the framework of GR?  
This would require a careful underlying choice of conceptually-clear general framework,
in the sense we discuss below.
{\bf Q2:} what exactly is the connection between such models and the underlying string/M theory?
More precisely, 
to what extent can the agreements or otherwise of predictions of such models 
with observations be taken as support or disagreement with such theories?
Here we concentrate on the first question
and make a comparative study of the general schemes that 
have been employed in the literature in order to construct the bulks which surround branes. 

Such a comparative study requires a sufficiently general common language.  We use that of 
\it p.d.e problems\normalfont, which consist of both the p.d.e system to be solved 
in some region of a manifold, and data prescribed on (portions of) the boundary of this region. 
For ($r$, $s$) a spacetime of dimension $n = r + s$ with $s$ time dimensions, we  
denote the problem involving the addition of an extra dimension by ($r$, $s$; $\epsilon$), 
where $\epsilon = 1$ if the new dimension is spacelike or $\epsilon = - 1$ if it is 
timelike.  This is the generalization of the GR Cauchy problem (CP) \cite{Cauchylit, HE, CBY, Wald} based on the 
arbitrary ($r$, $s$; $\epsilon$) generalization of 
the Arnowitt--Deser--Misner \cite{ADM} split of the metric (Sec 2.1) and hypersurface geometry (Sec 2.2).  Its simple 
signature-independent features are pointed out in Sec 2.3 and the crucial dependence of many of the harder features on 
the usual CP signatures $s = 0$, $\epsilon = -1$ is discussed in Sec 2.4.  As is well-known, 
the GR CP presupposes the existence of the data.  Thus one is in fact considering a two-step 
process, the other step being the construction of the data on the ($r$, $s$) manifold i.e the 
generalization of the GR initial value problem (IVP) \cite{Yorklit, MTW, Cook}.  This is discussed in Sec 3.  

Our framework permits a profitable look at a number of recent topics.  
The aim is to consider GR-based models containing thin matter sheets such as branes or domain walls.  
We shall compare two broad schemes that have been proposed to construct bulks: 
the (3, 1; 1) construction \cite{311lit, starapp, VW}
starting from information on a 

\noindent (3, 1) spacetime hypersurface (usually the brane) and 
the (4, 0; --1) construction \cite{401SS, 401N} starting 
by the construction of data on a (4, 0) spatial hypersurface.  
However, first we emphasize that one should grasp the fundamental arguments and results 
which before the specializing to the thin matter sheet models.  
A few ideas about the general unspecialized framework recently arose in the literature 
on (generalizations of) the Campbell--Magaard arbitrary-signature embedding theorem 
\cite{Campbell, Magaard, RTZ, ADLR} (see Sec 2.3).  However, 
we find it far more profitable instead to adopt the generalized GR CP--IVP point of view since this literature  
is by far more developed and thus a far greater source of well-documented pitfalls and carefully thought-out techniques 
which avoid them.  We identify the embedding step of the Campbell--Magaard theorem with the well-known signature-independent 
parts of the GR CP in Sec 2.3.  But the harder $s = 0$, $\epsilon = -1$ specific parts of the GR CP strongly suggest 
that (4, 0; --1) schemes should be favoured over (3, 1; 1) ones on very general grounds: well-posedness and causality (Sec 2.4).   
Furthermore, Magaard's data construction method \cite{Magaard} (Sec 3.1) does not compare favourably with York's data construction \cite{Yorklit} (Sec 3.2), 
and its application to $s = 1$, $\epsilon = 1$ has further conceptual difficulties.  In this light we look at the extent 
to which York's method is adaptable to $s = 1$, $\epsilon = 1$, and also consider the thin sandwich method \cite{BSW, TSC} 
in this context (Sec 3.3).

We then introduce thin matter sheets in Sec 4, and study the (3, 1; 1) schemes with thin matter 
sheets, recollecting the derivation of the junction conditions (Sec 4.1), and showing how the 
SMS formulation (Sec 4.2) may be reformulated in a large number of ways using 
geometrical identities that interchange which terms are present in the braneworld EFE's.  
This is illustrated by the formulation in Sec 4.3 which directly parallels the GR CP 
formulation and thus makes no explicit use of the Weyl term, and by formulations in Sec 4.4 in 
which the quadratic term has been re-expressed entirely in terms of derivatives off the brane.  
These formulations are used to clarify a number of aspects of the `Weyl problem', in particular 
to argue for the study of the full brane-bulk system.  Further aspects of such formulations are 
discussed in Sec 4.5.  The implications of Sec 2.4 for thin matter sheets, the heuristic use 
and limitations of the (3, 1) York method with thin matter sheets, and formulations of thin 
matter sheet thin sandwich conjectures are discussed in Sec 4.6.  

Having gathered arguments against using (3, 1; 1) schemes in Secs 2.4 and 3, we further consolidate 
this viewpoint by arguing in Sec 5 against a suggested virtue of (3, 1; 1) schemes \cite{sing1, sing2}: their use to remove singularities.  
In addition to using the framework of Sec 2, the corresponding study of geodesics 
is also used here, which presents further conceptual complications relevant to thin matter sheet models.  These 
follow from the 

\noindent (3, 1) geodesics not usually being among the (4, 1) geodesics.      

In Sec 6, we continue the study of the ($n$, 0; --1) scheme favoured by our arguments, 
in the presence of both thin and thick (i.e finitely thin) matter sheets.  
We begin in Sec 6.1 by providing a hierarchy of very difficult general thin 
matter sheet problems in which the main difficulties stem from low differentiability and 
details about the asymptotics.  Within this class of problems we identify how 
the far more specific scenarios currently studied emerge as more tractable cases, 
and thus identify which as-yet unjustified assumptions such studies entail.  
The IVP step, of use in the study of how braneworld black holes extend into the bulk (the `pancake' versus `cigar' debate \cite{311lit}), 
involves fewer of such assumptions.  Thus in this paper we restrict attention to the 
($n$, 0) data construction problem (Sec 6.2), which we apply to thin matter sheets in Sec 6.3 
and more straightforwardly to thick matter sheets in Sec 6.4.\normalfont.  Sec 7 discusses \bf Q2 \normalfont and contains the conclusions of this 
paper as regards \bf Q1\normalfont.

\section{Embeddings and the GR Cauchy Problem}

In this section, we give splittings of EFE's with respect to a hypersurface of codimension 1.  
This results in 2 systems of equations that 
are treated as constituting two separate 
steps. The first, embedding or evolution step,
will be dealt with in this section and the second, data construction step, will be presented in Sec 3.  
The embedding step includes the GR Cauchy problem (CP) for $s = 0$, $\epsilon = -1$.  
We work with general ($r$, $s$; $\epsilon$) since many applications of interest have $s = 1$ and $\epsilon = 1$.  
Our approach is to treat the general case as far as possible, 
while taking careful note of the many important roles played by the signatures $\epsilon$ and 
especially $s$ (whereas a change in the dimension $n$ plays only a mild role).      

An important point regarding the well-known analytic p.d.e system approach
employed here is the identification of the input to be prescribed and
the output to expect. 
With the advent of schemes to construct bulks \cite{311lit, 401SS, 401N, starapp, Gregory, Wesson03},
the old questions\footnote{Following from the original (3, 0; --1) work on GR by Lichnerowicz \cite{Lich} and (Choquet)-Bruhat \cite{B52, B56, Cauchylit} 
(see Sec 2.4 and 3.1-2), Wheeler \cite{Wheeler} asked questions about prescriptions, which stimulated the York and thin-sandwich work 
(in Secs 3.2 and 3.3 respectively).} regarding these  prescriptions have become relevant in a new context.

\subsection{Splitting up the Metric}

We begin by recalling the arbitrary ($r$, $s$; $\epsilon$) 
Arnowitt-Deser--Misner (ADM) \cite{ADM} type split 
w.r.t the ($r$, $s$) `initial' hypersurface $\Upsilon_0$ of the 
higher-dimensional metric $g_{CD}$\fn{In this paper we use 
capitals for the higher-dimensional indices involved in a split
and lower-case for the lower-dimensional indices.
$g_{AB}$, $ h_{ab}$; 
$\nabla_{C}$, $D_{c}$; 
${\cal R}\normalfont_{ABCD}$, $R_{abcd}$; 
${\cal G}_{AB}$, $G_{ab}$; 
${\cal T}_{CD}$, $T_{cd}$ are 
respectively the higher and lower dimensional 
metrics; 
covariant derivatives; 
Riemann tensors (and similarly for Ricci tensors and Ricci scalars); 
Einstein tensors; 
energy--momentum tensors.} 
along with the split of its inverse $g^{CD}$,
\be
\begin{array}{ll}
g_{CD} = \left( \begin{array}{ll} \mbox{ } \mbox{ } \mbox{ } \mbox{ } \beta_i\beta^i + \epsilon\alpha^2  & 
\beta_d \\ \mbox{ } \mbox{ } \mbox{ } \mbox{ } \beta_c &  h_{cd} \mbox{ } \mbox{ } \mbox{ } \mbox{ } \mbox{ } \end{array}\right)
\mbox{ }  , \mbox{  }  \mbox{  } 
g^{CD} =  \left( \begin{array}{ll}  \epsilon\frac{1}{\alpha^2} & 
-\epsilon\frac{\beta^d}{\alpha^2}
\\ -\epsilon\frac{\beta^{c}}{\alpha^2} & h^{cd} + \epsilon 
\frac{\beta^c\beta^d}{\alpha^2}\\ \end{array}\right)
\end{array}.
\label{ADMs}
\ee
The geometric interpretation of (and further notation for) 
the ($n$, 0, --1) and ($r$, 1, 1) splits discussed in this 
paper are exhibited in Fig. 1.  For the specific ADM case, 
the added dimension (of signature $\epsilon = -1$) is time, 
and the lower-dimensional hypersurface is 3-dimensional and 
entirely spatial: ($r$, $s$) = (3, 0).    The EFE's are 
then decomposed into 4 constraints and 6 evolution equations 
(Sec 2.3).  
The GR IVP and CP are defined w.r.t this split.  
The GR IVP is the solution of the 4 constraints [on some region $S_0$ of 
some (3, 0) initial hypersurface $\Sigma_0$] for initial data 
$(h_{ab}, K_{ab})$, where $K_{ab}$ is the extrinsic 
curvature
\be
K_{ab} = -\frac{1}{2\alpha}\delta_{\vec{\beta}}{h}_{ab},  
\label{ecd}
\ee
which is a measure of how much the hypersurface is bent in the surrounding spacetime.\fn{We use $\pounds_{\beta}$ to denote the Lie 
derivative w.r.t $\beta_i$, and a dot to denote the 
derivative in the direction perpendicular to the hypersurface.  The hypersurface derivative is given by 
$\delta_{\vec{\beta}} \equiv \frac{\pa}{\pa\lambda} - \pounds_{\beta}$ where ${\vec{\beta}}$ is the 4-vector $[\alpha, \beta_i]$.}
This data, together with the 6 evolution equations (for $\dot{K}_{ab}$) and the 
6 equations for $\dot{h}_{ab}$ which follow from (\ref{ecd}) constitute the GR CP.  
This is a (3, 0; --1) problem for the construction of the spacetime region in which $S_0$ is embedded.  

It is useful to juxtapose here another metric split, the arbitrary ($r$, $s$; $\epsilon$) KK type split  
\be
\begin{array}{ll}
g_{CD} =
\left(
\begin{array}{ll}
\epsilon\Phi^2 & \epsilon\Phi^2a_{d}\\ \epsilon\Phi^2a_{c} &  h_{cd} + 
\epsilon\Phi^2a_{c}a_{d}
\end{array}\right)
\mbox{  } , \mbox{  }  \mbox{  }
g^{CD} =
\left(
\begin{array}{ll}
a_{m}a^{m} + \epsilon\Phi^2 & - a^{d} \\ - a^{c} &  h^{cd} 
\end{array}\right)
\end{array}, 
\label{KKs}
\ee
so as to discuss some of the literature \cite{Wesson}.      
Note first that for KK theory itself, this split is (3, 1; 1), with $h_{cd}$, $a_{c}$, $\Phi$ 
held to be independent of the added dimension's coordinate $z$.  This so-called cylindricity 
condition implicitly permits $a_{c}$ to be interpreted as the classical 
electromagnetic potential.  More recent generalizations of this scheme, called 
`noncompact KK (NKK) theory' \cite{Wesson}, have involved 
large extra dimensions in place of cylindricity, by 
permitting $h_{cd}$, $a_{c}$ ,$\Phi$ to depend on $z$.  
In some simple instances, $a_{c}$ is held to be 0. 
But clearly for $a_c = 0$ and $\beta_c = 0$, the two 
splits (\ref{ADMs}) and (\ref{KKs}) are identical.\fn{This requires the 
identification $\alpha \leftrightarrow \Phi$. 
More generally, a KK split is the \it inverse \normalfont of an ADM split with the identifications 
$\alpha \leftrightarrow \frac{1}{\Phi}$, $\beta_i \leftrightarrow - a_i$.} \mbox{ }   
So there is already a vast literature on embeddings with a large extra dimension 
since this is what is usually treated in the GR CP.

In Sec 4, we reinterpret (3, 1; 1) methods such as that of SMS as \it z-dynamics \normalfont i.e ``dynamics"  
with the coordinate of the new dimension, $z$, as ``dynamical variable".\fn{We use $\lambda = z$
when the new dimension is spacelike and $\lambda = t$ when it is timelike.  We use one more set of apostrophes than is necessary to codify the nonstandard- 
or general-signature use of concepts that are usually considered in the context of a particular signature: dynamics, momentum, evolution, the Cauchy problem.}  
Note that such a ``dynamical" interpretation was of no avail in KK theory proper, 
for there $h_{cd}$ was taken to be independent of $z$.  
Note also that although a way of relating SMS's BEFE's and NKK has been pointed out \cite{PDL01}, 
we emphasize that it is far more useful as here to relate the steps leading to SMS's BEFE's to the GR CP and IVP literature, since this is 
by far more developed and thus a good source of ideas and theorems, on which the critical Sec 2.4 is based.

\subsection{A Full Account of Hypersurface Geometry}

Gauss' {\it outstanding theorem} ($R = 2K^2$ for a 2-sphere embedded in 
flat 3-space) already provides a good intuitive feel both of why matter squared 
terms arise in brane cosmology and of why intrinsic curvature singularities can in
principle be removed upon applying 
an embedding by being balanced out by extrinsic-curvature-squared terms.  
We work with the suitable generalization of this theorem: the ($n$ + 1)-dimensional geometry of projections 
w.r.t a ($r$, $s$) hypersurface.    
We use ${\cal O}_{...a...} = {\cal O}_{...A...}h^A_a$ for 
projections onto the ($r$, $s$) hypersurface $\Upsilon$, and 
${\cal O}_{...\perp...} = {\cal O}_{...A...}n^A$ for projections onto the normal.  
The three projections of the Riemann tensor \cite{Kucharlit} 
are respectively the Gauss, the Codazzi and the Ricci equations:  
\be
{\cal R}_{abcd} = R_{abcd} - 2\epsilon K_{a[c}K_{d]b},
\label{Gaussful}
\ee
\be
{\cal R}_{\perp abc} =  -2\epsilon D_{[c}K_{b]a},
\label{codful}
\ee
\be
{\cal R}_{\perp a\perp b} =  \frac{1}{\alpha}(\delta_{\vec{\beta}} K_{ab} - \epsilon D_bD_a\alpha) + {K_a}^cK_{cb}.
\label{thirdproj}
\ee
The Ricci equation (\ref{thirdproj}) is particularly important in Sec 4.3-4.  Contracting these equations gives  
\be
{\cal R}_{bd} - \epsilon {\cal R}_{\perp b\perp d} =  R_{bd} - \epsilon(KK_{bd} - {K_{b}}^{c}K_{cd}).
\label{contG}
\ee
\be
{\cal G}_{a\perp} = {\cal R}_{a\perp} = -{\epsilon}(D_b{K^b}_a - D_aK),
\label{Gpa}
\ee
\be
{\cal R}_{\perp\perp} =  \frac{\delta_{\vec{\beta}} K - \epsilon D^2\alpha}{\alpha} - 
K\circ K .
\label{tpcont}
\ee 
The Gauss equation may be contracted a second time to give (using $K \circ K \equiv K_{ij}K^{ij}$) 
\be
2{\cal G}_{\perp\perp} = 2{\cal R}_{\perp\perp} - \epsilon{\cal R} = - \epsilon R + K^2 - K\circ K 
\label{Gpp}.
\ee

The Ricci equation can be used in the contracted Gauss equation to obtain 
\be
\frac{1}{\alpha}(\delta_{\vec{\beta}} {K}_{ab} - \epsilon D_bD_a\alpha) - KK_{ab} +2{K_{a}}^cK_{bc} + \epsilon R_{ab} = \epsilon{\cal R}_{ab}.
\label{protoev}
\ee
The trace of this [a linear combination of (\ref{Gpp}) and (\ref{tpcont})] is:
\be
\frac{1}{\alpha}(\delta_{\vec{\beta}} {K} - \epsilon D^2\alpha) - K^2 + \epsilon R = \epsilon{\cal R}  - {\cal R}_{\perp\perp}
\label{prototrev}.
\ee
Another linear combination of (\ref{Gpp}) and (\ref{tpcont}) is
\be
\frac{\delta_{\vec{\beta}}K - \epsilon D^2\alpha}{\alpha} - \frac{K\circ K + K^2}{2} = \epsilon\frac{{\cal R}}{2}.
\ee
From this and (\ref{protoev}) it follows that 
\be
\frac{1}{\alpha}\left[\epsilon(\delta_{\vec{\beta}}K_{ab} - h_{ab}\delta_{\vec{\beta}}K) -D_bD_a\alpha + h_{ab}D^2\alpha\right]  
+ \epsilon
\left(
2{K_a}^cK_{bc} - KK_{ab} + \frac{K\circ K + K^2}{2}h_{ab}
\right)
+ G_{ab} = {\cal G}_{ab}
\label{gid}
\ee

\subsection{Simple Signature-Independent Results for Embeddings}

It is well-known that the general ($r$, $s$; $\epsilon$) ADM-type split of the EFE's 
${\cal G}_{AB} = {\cal T}_{AB}$ gives rise via (\ref{Gpp}) and (\ref{Gpa}) to the ($n$ + 1) Gauss and Codazzi 
constraints\footnote{In ADM's Hamiltonian language for ($r$, 0; --1)  
these are known as the Hamiltonian and momentum constraints respectively.} 
\be
K^2 - K\circ K  - \epsilon R = 2{\cal G}_{\perp\perp} = 2{\cal T}_{\perp\perp} 
\equiv 2\rho,
\label{gauss}
\ee
\be
-\epsilon(D_b{K^b}_a - D_aK) = {\cal G}_{a\perp} = {\cal T}_{a\perp} \equiv j_a,
\label{cod}
\ee
together with $n(n + 1)/2$ ``evolution equations" w.r.t the ``dynamical variable" $\lambda$ 
\be
\frac{1}{\alpha}(\delta_{\vec{\beta}} {K}_{ab} - \epsilon D_bD_a\alpha) - KK_{ab} +2{K_{a}}^cK_{bc} + \epsilon R_{ab} = 
\epsilon{\cal R}_{ab} = \epsilon
\left(
S_{ab} - \frac{S}{n - 1}h_{ab}
\right) - \frac{{\rho}}{n - 1}h_{ab},
\label{evK}
\ee
obtained via (\ref{protoev}).   
Here $\rho$, $j_a$ and $S_{ab} \equiv {\cal T}_{ab}$ (N.B $S \neq {\cal T}$) 
are general matter terms which are usually prescribed as functions 
of matter fields that are governed by usually-separate field equations. 
A useful equation is the trace of (\ref{evK})
\be
\delta_{\vec{\beta}} K = \alpha
\left(
K^2 - \epsilon R - \frac{n\rho + \epsilon S}{n - 1}
\right) 
+ \epsilon D^2\alpha =
\alpha
\left[
K\circ K + \frac{(n-2)\rho - \epsilon S}{n - 1}
\right] 
+ \epsilon D^2\alpha,
\label{trevK}
\ee
the second equality of which follows from the Gauss constraint (\ref{gauss}).  
 
Also applying to $K_{ab}$ the decomposition of symmetric 
second-rank tensors $A_{ab}$ into their trace $A$ and tracefree part, 
$A^{\mbox{\scriptsize T\normalsize}}_{ab} \equiv A_{ab} - \frac{A}{n}h_{ab}$, 
the equations (\ref{gauss}) and (\ref{cod}) and (\ref{trevK}) take on the useful forms 
\be
K^{\mbox{\scriptsize T\normalsize}}\circ K^{\mbox{\scriptsize T\normalsize}} 
- \frac{n - 1}{n}K^2 + \epsilon R + 2\rho = 0,
\label{Agauss}
\ee
\be
D_b {K^{\mbox{\scriptsize T\normalsize}b}}_a - \frac{n - 1}{n}D_aK + \epsilon j_a = 0,
\label{Acod}
\ee
\be
\frac{1}{\alpha}(\delta_{\vec{\beta}} K - \epsilon D^2\alpha) - \frac{K^2}{n} = 
K^{\mbox{\scriptsize T\normalsize}}\circ K^{\mbox{\scriptsize T\normalsize}} 
+ \frac{(n-2)\rho - \epsilon S}{n - 1}.
\label{protoRay}
\ee
   
We now consider the simpler results of the (3 ,0; --1) GR CP, 
most of which hold irrespective of both $\epsilon$ and $s$.   
First, there is the two-step nature of the problem for the whole EFE system.  

\noindent\bf ``Evolution" or embedding step: \normalfont given as data $(h_{ab}, K_{ab})$ together with 
coordinate functions $\rho$ and $j_a$, all on some portion $U_0$ of $\Upsilon_0$, such that the 
constraints are satisfied, one is to solve the evolution equations (\ref{ecd}) and (\ref{evK}) as 
a prescription for evaluating  $(h_{ab}, K_{ab})$ on some portion $U_{\lambda}$ of a nearby 
$\Upsilon_{\lambda}$ i.e for a small increment of the dynamical variable $\lambda$.

\noindent\bf ``initial value" or data construction step: \normalfont to construct such data by solving the 
constraints (see Sec 3).

Second, provided that the constraints hold on some ``initial'' $U_0$, 
they are guaranteed to hold throughout the embedding spacetime region. 
This is most quickly seen by virtue of energy-momentum conservation and the 
Bianchi identities \cite{ADM}, making use of the Einstein tensor  projection form of the constraints.  

Third, constraints come hand-in-hand with gauge freedoms.  
Thus $n + 1$ of the metric components are freely-specifiable. 
For many applications a choice of gauge is required.  To begin with we mostly use the 
\it normal coordinate gauge \normalfont ($\alpha = 1$, $\beta_i = 0$), and the 
\it harmonic gauge \normalfont \cite{Cauchylit, Wald} (coordinates $x^{A}$ such that 
$\nabla^2x^A = 0$), though we use more elaborate 
gauge choices later.  We stress that careful consideration of gauge choice 
can be an important ingredient toward producing prescriptions that are generally robust and physically 
clear \cite{Y78}.   
  
Fourth, by choice of a suitable gauge it can be ascertained by the Cauchy--Kovalevskaya 
(CK) theorem\fn{Cauchy--Kovalevskaya theorem:  
Suppose one has $\Delta$ unknowns $\Psi_{\Delta}$ which are 
functions of the ``dynamical variable" $\lambda$ and of $\gamma$ other independent variables $x_{\gamma}$.  
Suppose one is prescribed on some domain $M$ the $\Delta$ p.d.e's of order $k$ of form  
$\frac{  \pa^{(k)}\Psi_{\Delta}  }{  \pa \lambda^{(k)}  } = F_{\Delta}$ for the $F_{\Delta}$ 
functions of $\lambda$, $x_{\gamma}$, $\Psi_{\Delta}$ and of derivatives of 
$\Psi_{\Delta}$ up to ($k$ - 1)th order w.r.t $\lambda$, where 
furthermore in $M$ the $F_{\Delta}$ are \it analytic functions \normalfont of their arguments.  
Suppose one is further prescribed \it analytic data \normalfont 
$\Psi_{\Delta}(0, x_{\gamma}) = {}^{(0)}d_{\delta}(x_{\gamma})$, ... , 
$\frac{  \pa^{(k-1)}\Phi_{\Delta}  }{  \pa \lambda^{(k-1)}  }(0, x_{\gamma}) = 
{ }^{(k -1)}d_{\Delta}(x_{\gamma})$ on some piece $U$ of a 
\it nowhere-characteristic surface\normalfont. 
Then this problem \it locally \normalfont possesses precisely one \it analytic \normalfont solution.  
Note 1) this is the most basic theorem for p.d.e's. 
2) It depends entirely on the functions being analytic, which seriously limits its applicability.  
3) It is a very general theorem in the sense that it holds for all types of p.d.e's.  
In this paper, this is reflected by the signature (principal symbol)-independence 
of the applications and by the irrelevance of what analytic function $F_{\Delta}$ is.} \cite{CH} 
for admissible analytic data that there locally exists a unique solution to the 
``evolution equations" (\ref{evK}).

For this and further applications in the GR CP, the harmonic gauge is usually chosen.  
In the ($n$, 0; --1) GR CP, the use instead of the normal 
gauge is discouraged because in practice it typically breaks down quickly.  
To see this recall that 
in normal gauge equation (\ref{protoRay}) may be written as the normal Raychaudhuri equation 
\be
\dot{K} - \frac{K^2}{n} = 
K^{\mbox{\scriptsize T\normalsize}} \circ K^{\mbox{\scriptsize T\normalsize}} + {\cal R}_{\perp\perp} \geq 0,  
\label{normRay}
\ee 
where the inequality follows from the definition of the strong energy condition (SEC): 
\be
\mbox{The general Raychaudhuri equation term } {\cal R}_{AB}u^Au^B \geq 0 \mbox{ } \mbox{ } \forall \mbox{ unit timelike } 
u^A,
\label{protoSEC}
\ee
by picking $u^A = n^A = t^A$.  To make (\ref{protoSEC}) manifestly an energy condition, 
one uses the EFE's to obtain 
\be
{\cal T}_{AB}u^Au^B \geq -\frac{{\cal T}}{n - 1} \mbox{ } \mbox{ } \forall \mbox{ unit timelike } u^A.
\label{SEC}
\ee
Integrating the differential inequality in (\ref{normRay}) shows that 
if $K_0 \geq 0$, $|K| \longrightarrow \infty$ within a finite range of the parameter 
$\pi = \frac{n}{|K_0|}$ along $t^A$. This blowup is by definition a \it caustic \normalfont 
and signifies a breakdown of the normal coordinates. However, for the ($r$, 1; 1) 
embedding $n^A = z^A$ is spacelike. 
There is then no good reason for ${\cal R}_{\perp\perp}$ 
to be always positive, so caustics may be less likely to form.  
The issue of the longevity of normal coordinates is of interest 
because they have hitherto been used for some of the applications below. 
Other applications below 
fix $\beta_i = 0$ alone.

An embedding result we discuss below is 
the Campbell--Magaard (CM) theorem, according to which any $n$-space 
can be surrounded by $n$ + 1 dimensional vacuum space, where the spaces in question are analytic.  
This is closely related to the above split of the EFE's; in particular it has an embedding 
step which we consider in this section and a data construction step, `Magaard's method', 
which we consider in Sec 3.1. The embedding step of the CM theorem is simply the 
result above based on the CK theorem, which is well-known for the GR CP, 
and follows for the other ($r$, $s$; $\epsilon$) embedding cases from the well-known 
insensitivity of the CK theorem to ($r$, $s$; $\epsilon$).  
Strictly, the CM result is the vacuum case (${\cal T}_{AB} = 0$) 
but the proof may easily be extended to any functional form of ${\cal T}_{AB}$ \cite{ADLR} 
since the use throughout of the CK theorem is totally insensitive to 
changes in such subleading order terms (contributions to $F_{\Delta}$).  
There have been suggestions 
that such \cite{Wesson03} or similar \cite{311lit} (3, 1; 1) 
methods could be used for the construction of bulks.   
However, even though the above results in the ($n$, 0; 1) case of the GR CP are 
protected and extended by a host of further theorems, 
they do not carry over to the (3, 1; 1) case, as we argue in the next section. 
There are also problems with (3, 1) data construction, as we shall see in Sec 3.

\subsection{The Breakdown of the Cauchy Problem -- Embedding Analogy}

In this section we argue by further comparison with the GR CP 
that the popular practise of building bulks by $z$-dynamics is in general unsatisfactory.  
For, in the GR CP a large number of important results \cite{B52, Leray, B56, Cauchylit, HE, HKM, CBY, RF, Klainerman} 
for the GR CP turn out to be entirely dependent on the lower-dimensional signature $s = 0$.  
In other words, the choice of methods which properly respect the difference between 
space and time is absolutely crucial.

To have a sensible problem in mathematical physics, one requires not just local existence and 
uniqueness of solutions but rather \it well-posedness\normalfont.   
Whereas well-posedness always includes local existence and uniqueness, 
it also always includes continuous dependence of the data, without which an arbitrarily small change in the data could 
cause an arbitrarily large immediate\fn{We emphasize that the continuous dependence in question (see e.g page 229 of \cite{CH}) 
is an immediate and discontinuous change in behaviour, not an issue 
of chaos or unwanted growing modes (though well-posedness often also bounds the growth of such modes \cite{RF}).}
 change in the evolution \cite{CH} i.e there is no guarantee of physical predictability from such a problem.  
For a hyperbolic-type system one further requires a domain of dependence (DOD) property to enforce a 
sensible notion of causality.  If the data is known only on a closed achronal set $S$ 

\noindent (that is, 
a piece of a spacelike hypersurface), then the evolution can only be predicted within a region 
${\cal D}^+(S)$ (the future DOD), defined as the set of points such that all past-inextendible 
causal curves through each point [represented by the $\gamma$'s leading to the point $p$ in 
Fig. 4a)] intersect $S$.  In our view causality can effectively be studied 
only in settings where the dynamical variable is time.  One reason for this is that given 
information on an arbitrarily-thin (3, 1) hypersurface,
the (4, 1)-dimensional DOD is negligible because
of this thinness (see Fig. 3).  Other reasons are discussed below.   

Now we note that the Cauchy--Kovalevskaya theorem has no control whatsoever over these last two properties for 
the analytic functions.  There are furthermore two reasons why the analytic functions are 
undesirable.  First, the analytic functions fail to be appropriate for any causal study since they are rigid 
(a change in a small region of an analytic data set induces 
an unwanted change of the entire data set).  Without them one absolutely cannot have anything 
like the extremally general (and in particular signature-insensitive) Cauchy--Kovalevskaya theorem.
Second, realistic\fn{Whereas Hawking and Ellis \cite{HE} argue that the choice of function space used 
to model nature does not matter since it is not experimentally-determinable and 
in any case is only an approximation due to QM,  we would refine this argument to say that 
the difference between the analytic functions and other function spaces is important because of the rigidity problem.  
Beyond that, we do not know if the particular function spaces  
used to prove rigorous theorems about the EFE's and about the extendibility of spacetime \cite{Clarke} (of relevance to our Sec 5) 
may be substituted in these applications by approximations based on other function spaces.  
In this case, it may be a mere matter of convenience: one wants to use whichever sufficiently 
general function spaces accessibly give rise to the strongest possible theorems.    }
and (or) interesting situations seldom involve analytic or smooth 
functions.  For example astrophysical bodies have boundaries, thin matter sheets are discontinuous 
and the approach to spacetime singularities can be of low differentiability.  

Bruhat showed that the GR CP is well-posed in 1952 (\cite{B52}) and 1956 (\cite{B56}),  
by making use of how the

\noindent ($n$, 0; --1) EFE system (\ref{gauss}--\ref{evK}) when cast in harmonic coordinates 
is of the correct quasilinear hyperbolic form 
${\cal L}^{\Gamma\Delta}(x,\Phi_{\Delta},\nabla_{\Delta}\Phi_{\Gamma})\nabla_{\Gamma}\nabla_{\Delta}\Phi_{\Sigma} = {\cal F}_{\Sigma}(x,\Phi_{\Delta},\nabla_{\Delta}\Phi_{\Gamma})$  
(where ${\cal L}^{\Gamma\Delta}$ is a Lorentzian metric; both this and the function ${\cal F}_{\Sigma}$ are smooth) to enable use of Leray's theorem \cite{Leray, Wald}, which 
guarantees local existence and uniqueness, and furthermore continuous dependence on the initial data and the DOD property.  

The above proofs of the four well-posedness 
criteria for the GR CP are now usually done 
using \it Sobolev spaces \normalfont \cite{Wald, HE, HKM, CBY, Clarke, Klainerman}. 
These are appropriate because of the hyperbolic character of the Einstein evolution equations.  
At first they lend themselves to less involved proofs than Bruhat's, although 
if one seeks yet stronger results the functional analysis becomes extremally unpleasant.  
To recollect simply why Sobolev spaces are appropriate \cite{Wald}, it suffices to consider a Klein--Gordon 
scalar in Minkowski spacetime.    
Given data on a bounded region $S_0$ of a spacelike hypersurface $\Sigma_0$, one can draw the 
future DOD ${\cal D}^+(S_0)$ [Fig. 4a)] which is the region affected solely by this data due 
to the finite propagation speed of light.  One can then consider the values of $\psi$ and its 
first derivatives on $S_t = {\cal D}^+(S_0) \cap \Sigma_t, t > 0$.  Then by Gauss's (divergence) 
theorem and energy-momentum conservation, 
\be
\int_{S_0}{\cal T}_{AC}t^{A}t^{C} + 
\int_{B}{\cal T}_{AC}l^{A}t^{C} =
\int_{S_t}{\cal T}_{AC}t^{A}t^{C},
\ee
and the second term $\geq 0$ provided that the matter obeys the dominant energy condition (DEC) 
\be
-{{\cal T}^A}_Cu^C \mbox{ is a future-directed timelike or null vector } \mbox{ }\forall \mbox{ } \mbox{future-directed timelike } u^A
\ee
and that $t^{A}$ is timelike.  Then the definition of the energy-momentum tensor gives
\be
\int_{S_t}((\dot{\psi})^2 + (\nabla\psi)^2 + m^2\psi^2) \leq
\int_{S_0}((\dot{\psi})^2 + (\nabla\psi)^2 + m^2\psi^2)
\label{sums}
\ee
Because each integrand is the sum of squares (which are necessarily positive), this means that control over the data on $S_0$  
gives control of the solution on $S_t$.  
The idea of a Sobolev norm is a generalization of these last two `energy' integrals \cite{Clarke}.  
For later use, the Sobolev class $H^w$ has a bounded norm of 
this type including up to $w$th order derivatives.

Whereas the GR CP comes with important well-posedness theorems 
\cite{Wald, HE, CBY}, the ($r$, 1; 1) 
problem is instead a complicated instance of a \it sideways Cauchy 
problem\normalfont, a category of problem for which the possibility of such well-posedness  
theorems remains notably undeveloped.\fn{The few simple results known for flat spacetime 
sideways wave equation problems are collected in \cite{AS}; these results might 
serve as a starting point for the study of the much more complicated nonlinear sideways ``GR CP" 
system.  So little is known about ultrahyperbolic equations that we deem it not sensible 
to talk about ($r$, $s$; $\epsilon$) procedures  at present for $s > 1$ or $s = 1$, 
$\epsilon = - 1$.  In this paper we argue that even $s = 1$ $\epsilon = 1$ is hard enough!} 
There is quite simply no established way to proceed.      
There is no known sideways ($s = 1$) analogue of Leray's theorem. 
There is no known notion of what function spaces might be appropriate for sideways Cauchy problems.  
We can explain however why the above introduction of Sobolev 
spaces does not carry over to sideways Cauchy problems.  
First there is no sideways notion of DOD to make the construction.  
Second even if we assume the higher-dimensional DEC holds, it would not give an inequality because the 
perpendicular vector $z^A$ in now spacelike [Fig. 4b)].  Third, we obtain a difference of squares 
rather than the sums in (\ref{sums}), so the equivalent of the energy method's use of Sobolev norms 
is simply of no use to control the ``evolution" given the data.

\section{Comparison of Methods of Data Construction}

Several methods have been proposed for the data construction step, including the elimination and
conformal methods.  The former methods are intuitive in that the prescribed 
quantities are all physical, but have a number of 
undesirable mathematical features. In contrast, Lichnerowicz's conformal 
method \cite{Lich} is counterintuitive,
since some of the prescribed 
quantities are unphysical, but 
exploits the mathematical properties of the
constraint system 
in order to decouple it. Bruhat's argument \cite{B56} was in favour of the latter, however,
some of her 
criticisms need to be elaborated upon to cover Magaard's elimination method \cite{Magaard}.  
Sec 3.1 covers this question and provides further criticisms in the $s = 1$ case. In Sec 3.2, we discuss 
the desirable features of York's development of the conformal method \cite{Yorklit, MTW}, paying careful attention to the
signature-dependent ones. The thin sandwich and Hamiltonian methods are briefly discussed in Sec 3.3.

\subsection{Elimination methods}

An example of this method of data construction 
is employed in the last part of the CM result 
(``Magaard's method'' \cite{Magaard, ADLR}). 
We discuss and compare this method with the standard method 
used for the GR IVP. Magaard's method (1963) treats the lower-dimensional metric $h_{ab}$ 
as a known.  Here we use a more transparent notation than Magaard's, in which the Gauss 
constraint (\ref{gauss}) takes the form
\be
G^{abcd}K_{ab}K_{cd} + \epsilon R + 2\rho = 0
\ee
where $G^{abcd} = h^{ac}h^{bd} - h^{ab}h^{cd}$ is the (undensitized) inverse of 
DeWitt's supermetric \cite{DeWitt}.  This is then split w.r.t some coordinate $x_1$. 
Then $K_{11}$ can be isolated as 
the solution of the linear equation\fn{The Gauss constraint is linear, not quadratic, in $K_{11}$ by an obvious antisymmetry 
of the inverse supermetric.  This split uses normal coordinates so $h_{1a} = 0$ and $h_{11} = 1$.}
\be
2K_{11}G^{11uw}K_{uw} + G^{1u1w}K_{1u}K_{1w} + G^{uwxy}K_{uw}K_{xy} = - (\epsilon R + 2\rho)
\ee
(where $u$, $w$, $x$, $y$ $\neq 1$), provided that it is possible to divide by $Z \equiv G^{11uw}K_{uw}$.  Thus $K_{11}$ is eliminated  
from the Codazzi constraint (\ref{cod}), which is then treated as a p.d.e system for unknowns 
${\cal U} \equiv \{K_{1w}$, some $K_{uv}$ component denoted $E$\}.  
On the face of it, this  system satisfies the criteria for the CK theorem 
if one treats ${\cal U}^{\prime} \equiv$ \{all the components of $K_{ij}$ bar $K_{11}$, $K_{1u}$ and $E$\} 
as known functions on $x_1$ and provided that the p.d.e's coefficients and the data are analytic. So a unique solution 
exists.  The (generalized) CM result \cite{Magaard, ADLR} groups this and the local existence of a unique ``evolution" to 
form the statement that a ($r$, $s$) spacetime with prescribed analytic metric $h_{ab}$ may be embedded with an extra space 
or time dimension for  any analytic functional form of the energy-momentum tensor.  This statement suggests an incredibly 
rich collection of embeddings exist.  However, we must point out that this is itself one of many difficulties 
associated with the CM result and its applicability, that we describe below.          

Bruhat had already considered the above method in 1956 \cite{B56} in the case of the GR IVP and pointed out a 
limitation on its validity.  We find that her argument can be made for any signature or dimension: $Z$ must be a function 
of at least one unknown, $E$,  so we have no control over whether it is zero.  Therefore we cannot guarantee the validity 
of the elimination procedure for $K_{11}$ from the Gauss constraint.  Although Magaard \cite{Magaard} finds a route round this problem, we 
find that this leads to two prices to pay later on .  He starts with prescribed `data for the data' i.e values of ${\cal U}$ on 
some $(n - 1)$-dimensional set ${\cal X}_1 = \{x_1 = 0\}$.  The `data for the data' is validly picked so that $Z \neq 0$ on ${\cal X}_1$, 
whereupon by continuity there is a thin region $0 \leq x_1 < \eta$ within which it is guaranteed that $Z \neq 0$.  Thus Magaard's 
procedure produces strips of data to be used in the embedding step.  
The first price to pay is that in general we cannot expect to be able to patch such 
strips together to make extended patches of data.  For, since the strip construction ends where $Z$ picks up a zero for some $x_1 = \eta$, 
while restarting the  procedure with $x_1 = \eta$ in place of $x_1 = 0$ is valid, the two data strips thus produced will have a discontinuity across $x_1 = \eta$.  
So what one produces is a collection of strips belonging to different possible global data sets.  
The evolution  of each of these strips would produce pieces of different higher-dimensional manifolds.  
So one is \it not \normalfont in fact saying that an empty ($n$ + 1)-dimensional manifold surrounds any $n$-dimensional 
manifold.  Rather one is saying that any $n$-dimensional manifold can be cut up in an infinite number of ways 
(choices of the $x_1$ coordinate) into many pieces, each of which can be separately bent in an infinite 
number of ways (corresponding to the freedom in choosing the components of $K_{ab}$ in ${\cal U}^{\prime}$ on each set of `data for the data'), 
and for each of these bent regions we can locally find a surrounding ($n$ + 1)-dimensional manifold for every possible analytic function form of 
${\cal T}_{AB}$ (corresponding to the generalizability of the CM result).  
This apparent excess richness raises the question of physical significance of such a construction.

Magaard's method does not state enough assumptions to make it rigorous.
First, how far does `data for the data' extend along $x_1 = 0$?  
Clearly the topology of the $n$-dimensional 
manifold is an important input, for if it is not compact without boundary, 
there is a missing boundary or asymptotics 
prescription required.
Also the topology of the $x_1 = 0$ set itself has not been brought into 
consideration (for example continuity is not guaranteed were this to contain loops).  
 
Secondly, we specifically consider the ($n$, 0; --1) and ($r$, 1; 1) problems as separate cases, since we 
find that

\noindent there are implicit ways in which signature plays a crucial role even for the CM result.  For 
($n$, 0; --1) one might 
worry that the data construction for a strip is flawed because the data problem in question is elliptic and hence naively 
requires a global treatment.  However, we are able to dispel this worry once we treat York's method below.  For ($r$, 1; 1), 
the use of the strip $0 \leq x_1 < \eta$   as ``evolutionary" data is generally invalidated by 
the information leak construction in 
Fig. 5, unless one has had the luck to construct a full global data set.  
This would, however, require the building of the data encountering 1) no zeros 
of $Z$ (which is thus the second price to pay), 2) no asymptotic problems. 
Thus in general there is a severe problem with building (4, 1) spacetimes 
from (3, 1) ones using the CM method.

Thirdly, Magaard's method lacks any $n$-dimensional general 
covariance since it involves the choice of a coordinate $x_1$ and the elimination 
of a ${11}$-component of a tensor.  

\subsection{Conformal methods}

The arguments of Lichnerowicz \cite{Lich} and Bruhat \cite{B56} led to the GR IVP 
taking a very different route from the above sort of brute-force elimination methods.      
The conformal method of Lichnerowicz and York is instead preferred.  
This leads us to ask to what extent conformal methods can be adapted both to heuristic 
and to general constructions for e.g ($r$, 1; $\epsilon$) data.  
Therefore our treatment below is as far as possible for the general ($r$, $s$; $\epsilon$) case.  

In the conformal method, one chooses to treat $h_{ij}$ as a known metric which is moreover not 
the physical metric but rather only conformally-related to it by 
\be
\tilde{h}_{ij} = \phi^{\eta}h_{ij}.  
\ee
We work in terms of $K^{\mbox{\scriptsize T\normalsize}}_{ij}$ and $K$, 
and permit $K^{\mbox{\scriptsize T\normalsize}ij}$, $j^i$ and $\rho$ to conformally transform according to 
\be
\tilde{K}^{\mbox{\scriptsize T\normalsize}ij} 
= \phi^{\zeta - 2\eta}K^{\mbox{\scriptsize T\normalsize}ij} \mbox{ } , \mbox{ } \mbox{ } \tilde{j}^i 
= \phi^{\xi} \mbox{ } , \mbox{ } \mbox{ } \tilde{\rho} = \phi^{\omega}\rho, 
\ee
whilst crucially demanding the constant mean curvature (CMC) condition   
\be
K = \mbox{hypersurface constant}
\ee
holds and is conformally-invariant.  
One then demands that the (raised) Codazzi constraint (\ref{Acod}) is to be conformally-invariant. Since 
\be
\tilde{D}_a\tilde{K}^{\mbox{\scriptsize T\normalsize}ab} = \phi^{\zeta - 2\eta}
\left[
D_aK^{\mbox{\scriptsize T\normalsize}ab} + 
\left(
\zeta - 2\eta + \eta\frac{n + 2}{2}
\right)K^{\mbox{\scriptsize T\normalsize}ab}\frac{D_a\phi}{\phi}
\right]
\ee
one requires that 
\be
-\eta\frac{n + 2}{2} \mbox{ } = \zeta - 2\eta = \xi.
\label{twiddle}
\ee
Furthermore one demands that the conformally-transformed Gauss equation 
\be
R - \eta(n - 1)\frac{D^2\phi}{\phi} + \eta(n - 1)\left( 1 - \eta\frac{n - 2}{4}\right)\frac{|D\phi|^2}{\phi^2} 
+ \epsilon(M\phi^{2\zeta - \eta} - 
\mu^2\phi^{\eta} + 2\rho\phi^{\omega + \eta}) = 0
\label{prolich}
\ee
(where $M = K^{\mbox{\scriptsize T\normalsize}} \circ K^{\mbox{\scriptsize T\normalsize}}$ and $\mu$ 
is some ``York extrinsic dynamical variable" proportional to $K$) 
is simplified by being made to contain no $|D\phi|^2$ term, so that 
$\eta = \frac{4}{n - 2}$, $\zeta = - 2$ and $\xi = -2\frac{n + 2}{n - 2}$ 
[by making use of (\ref{twiddle})].  Now, regardless of ($r$, $s$; $\epsilon$), 
provided that the ($n$ + 1)-dimensional DEC is to be preserved by the
conformal transformation,  $\rho^2$ must conformally-transform like $j^aj_a$, implying 
$\omega = -2\frac{n + 1}{n - 2}$.
Then (\ref{prolich}) becomes the ($r$, $s$; $\epsilon$) version\fn{The main feature due to $s$ is hidden in $D^2$.  In the specific case of $s = 0$, 
$D^2$ is the elliptic Laplacian operator denoted by   
$\triangle$.}  of the
Lichnerowicz equation
\be
D^2\phi = -\frac{\epsilon}{4}\frac{n - 2}{n - 1}\phi
\left( 
-\epsilon R - M\phi^{-4\frac{n - 1}{n - 2} } + \mu^2\phi^{\frac{4}{n - 2}} - 2\rho\phi^{-2\frac{n - 1}{n - 2}}
\right)
\label{ndlich}
\ee
for the conformal factor $\phi$.  We observe that a number of the very attractive 
features of the York method are ($r$, $s$; $\epsilon$)-independent.  First, it decouples the 
solution of the Codazzi constraint from the solution of the Gauss constraint.
The former proceeds by a traceless-transverse (TT)--traceless-longitudinal (TL) 
splitting \cite{Yorklit}
\be
K^{\mbox{\scriptsize T\normalsize}ij} = K^{\mbox{\scriptsize TT\normalsize}ij} 
+ K^{\mbox{\scriptsize TL\normalsize}ij} \mbox{ } , \mbox{ } \mbox{ } 
D_iK^{\mbox{\scriptsize TT\normalsize}ij} \equiv 0 \mbox{ } , \mbox{ } \mbox{ } 
K^{\mbox{\scriptsize TL\normalsize}ij} = 2\left(D^{(i}W^{j)} - \frac{1}{n}h^{ij}D_cW^c\right)
\ee
(for some vector potential $W^c$), which along with the trace-tracefree split is 
conformally-invariant and thus 
 unaffected by the solution of the latter, which has become the 
p.d.e (\ref{ndlich}) for the conformal factor.  The simplest
case is 
$K^{\mbox{\scriptsize T\normalsize}ij} = K^{\mbox{\scriptsize TT\normalsize}ij}$, $j^a  = 0$ which at most 
requires solution of a first-order linear equation.  

Second, the choice in using the scale-scalefree decomposition of the metric and the trace-tracefree 
and TT--TL decompositions of $K_{ij}$ are all decompositions into irreducibles and thus $n$-dimensionally 
generally-covariant choices, a decided advantage over the Magaard method.  

Third, one can attempt to preserve the CMC condition away from the `initial' 
hypersurface $\Upsilon_0$, which can be done provided that the equation 
\be
D^2\alpha + \alpha\left(\epsilon K^2 - R -\frac{\epsilon n\rho + S}{n - 1}\right) = \mbox{hypersurface constant}  
\ee
[from the first form of (\ref{trevK})] can be solved for the lapse $\alpha$.  This choice of lapse is the CMC slicing gauge.  
This deliberately defocusses geodesics, 
thus by definition preventing the breakdown of the coordinate system due to 
caustic formation \cite{Lich, Y78}, and so enhancing the practical longevity of the evolution. 
However, York's method is not absolutely general, for some spacetimes have 
no CMC slice to identify with $\Upsilon_0$ in the first place, 
and in others the CMC slicing cannot be maintained to cover the whole spacetime.  
Furthermore, these results depend on the asymptotics assumed.\fn{It was these 
caveats, along with the reliance on $s = 0$ of York's method and restrictions on the values 
$R$ can take for some subcases of the Lichnerowicz equation, that made us suspicious 
of the supposed generality of the Magaard method.}  
However, in the usual (3, 0; --1) case, this method (and its variants) 
is widely accepted as a practical method by the numerical relativity community, 
for example in the study of colliding compact astrophysical objects \cite{Y79book, Cook, BSrev}.  
Of importance for the viability of this method in this application,
recent evidence suggests that problems involving gravity wave emission can be treated thus \cite{Gowdy}.

However, some of the other attractive features used to gain control in the York method are only known to occur for the $s = 0$ case 
and have  mainly been studied in depth for the compact without boundary and asymptotically-flat cases of (3, 0) data construction.  
First, the study of the usual (3, 0; --1) Lichnerowicz equation has been based on its ellipticity 
(and elliptic methods are absolutely not generalizable to the 
`hyperbolic' equations\fn{Unlike ellipticity, hyperbolicity is actually difficult to pin down for nonlinear systems \cite{RF}.}).  
The (3, 0; --1) Lichnerowicz equation is well-studied, including with most fundamental matter fields \cite{IOY}, 
and for Sobolev spaces matching those then used in the GR CP \cite{B76, CBY}.  
Second, the study of the CMC slicing equation has been based on its ellipticity, 
so the theory of existence of CMC slicings \cite{Y78, CMClit} is likely to be 
signature-dependent.  
Third, the method depends on the asymptotics used and has only been studied in the (3, 0) 
compact without boundary \cite{cwb, CBY} and asymptotically flat \cite{Cantor, CBY} cases.  
Fourth, for $s = 0$ there is the useful property that certain local data patches can be proved to suffice for the treatment of an astrophysical problem, 
by building on the notion of DOD.  As mentioned,
this sort of technique is also applicable to protect pieces of local data obtained by 
Magaard's method on a ($r$, 0) hypersurface. 

Investigation of whether the $s = 1$ `wave Lichnerowicz equation' has good existence and uniqueness properties could be interesting.  
The natural setting for this is as a Cauchy problem (for even the 2-dimensional wave equation is ill-posed as a Dirichlet problem).  
Assuming that there exists a Cauchy surface in the (3, 1) spacetime sense, one 
can attempt forward and backward evolution to produce a global data set (assuming also that the 
decoupled procedure for finding ${\cal K}^{\mbox{\scriptsize T\normalsize}}_{\Gamma\Delta}$ also yields a global solution). 
One must remember however that the next stage is still to be a sideways Cauchy problem.  
Whilst this suggested procedure for the data could conceivably produce global data sets 
in some subcases and thus avoid the information leak problem, the other difficulties we 
described in Sec 2.4 remain. Therefore in our view the `wave Lichnerowicz equation' 
is unlikely to be suitable as a general method. We favour instead the (4, 0; --1) approach in Sec 6.  That 
said, the `wave Lichnerowicz equation' may still serve as the basis of a useful heuristic method.  
We consider this in Sec 4.7.

\subsection{Other methods and formulations}

We finally discuss the thin sandwich method and why we do not 
provide an ($r$, $s$; $\epsilon$) Hamiltonian formulation.  

In GR the lapse may be \it algebraically \normalfont eliminated 
from the action by use of its own variational equation to form a 
reduced Baierlein--Sharp--Wheeler action \cite{BSW}.
Suppose that ($h_{ij}$, $\dot{h}_{ij}$) is prescribed on a hypersurface
along with  $\rho$ and $j_i$.
The thin sandwich conjecture \cite{Wheeler} is that in
reasonably general circumstances there exists a unique solution to
solving the variational equation for the shift
(this equation replaces the Codazzi constraint), as a 
differential equation for the shift itself.
Once this is done, the lapse may be reconstructed from 
the equation used to eliminate it, and the
extrinsic curvature then follows from its definition. 
The lapse may then be reconstructed from the equation used to eliminate it, and the
extrinsic curvature then follows from its definition (\ref{ecd}).  

We note that this procedure is ($r$, $s$; $\epsilon$)-independent.
Thus we pose the conjecture for general ($r$, $s$; $\epsilon$).  
This is also considered for thin matter sheets in Sec 4.7.
To date we are only aware of (3, 0; --1) papers \cite{TSC} 
treating this conjecture, which contain some 
favourable and unfavourable results
obtained mostly by use of elliptic methods.
In fact, there are two variants of the conjecture: 
i) the thick sandwich conjecture in which lower-dimensional metrics 
are prescribed on two nearby hypersurfaces. 
ii) The thin sandwich conjecture proper, as stated above.

A (3, 1; 1) Hamiltonian formulation is not provided  
because although it 
is insensitive to $\epsilon$ \cite{Kucharlit}, it is sensitive to $s$ 
through the intimate involvement of the space of geometries.  
Whilst the usual Hamiltonian formulation is based on the space of Riemannian 
geometries (superspace) \cite{Wheeler, DeWitt}, the space of semi-Riemannian geometries 
is reported to be not even Hausdorff \cite{Haus}.  

\section{Application of (r , 1 ; 1) Methods to Thin Matter Sheets} 

First, we present a simple instance common in the higher-dimensional literature 
to which Sec 2.1-3 is applicable.  The \it warpfactor \normalfont split \cite{stdw2} 
\be
g_{CD} = 
\left(
\begin{array}{ll}
\Phi^2(x^{\Pi}, z) & 0 \\ 0 & W(x^{\Pi}, z)f_{\Gamma\Delta}(x^{\Pi})
\end{array}
\right)
\mbox{ } , \mbox{ } \mbox{ } W(x^{\Pi}, 0) = 1
\label{wan}
\ee
is a simple subcase of the $z$-dynamics scheme, in which the metric is allowed to 
$z$-evolve only in its scale, away from the $z = 0$ hypersurface where it is taken to be known.   
Then (\ref{ecd}) leads to $\pa f^{1/2}/\pa z = -f^{1/2}\Phi {\cal K}$ which gives the equation 
\be
\frac{\pa lnW}{\pa z} = -\Phi {\cal K}
\label{wftr}
\ee
for the warpfactor. For example, using the ansatz ${\cal K}_{\Gamma\Delta} = Cf_{\Gamma\Delta}$ 
in (\ref{wftr}) gives an exponential Randall--Sundrum type warpfactor.  Whereas 
the split (\ref{wan}) does not cover very many cases, our scheme exhibits generalizations for it: 
to permit the whole metric to evolve and to recognize the gauge freedom in 
$\beta_{\Gamma}$, which should ideally be used to separate coordinate effects 
from true physics in the spirit of \cite{Y78}.  

Full, overtly $z$-dynamical schemes are used in particular examples for domain walls \cite{Gregory}, 
braneworld black holes (such as for the pancake or cigar bulk horizon shape 
problem \cite{311lit}) and for braneworld stars \cite{starapp, VW}. Whereas in the 
Randall--Sundrum model the higher and lower-dimensional cosmological constants balance out 
leaving vacuum (Minkowski) on the brane, more generally a brane would consist of a thin sheet 
of matter -- a junction.  We study such [Shiromizu--Maeda--Sasaki (SMS)-type] 
braneworlds below,\fn{This section presents the material announced in \cite{ATlett} 
in a more advanced form and provides the derivations of the results along with a number of 
additional points and further new results.} starting first however with a careful 
recollection of where the underlying Israel junction conditions come from.   

\subsection{Origin of the Junction Conditions}

Assume that we have a ($D$ -- 1)-dimensional thin matter sheet in a $D$-dimensional bulk.  
In all cases considered, the bulk's extra dimension is spatial ($\epsilon = 1$).  
Our discussion follows \cite{MTW} most closely whilst keeping the unraised index positions of 
Israel's original work \cite{jns}.

Whereas the requirement of well-defined geometry dictates that the metric is 
continuous across the thin matter sheet yielding the junction condition (j.c)
\be
[f_{\Gamma\Delta}]^+_- \equiv f_{\Gamma\Delta}^+ - f_{\Gamma\Delta}^- = 0,
\label{jcf}
\ee
discontinuities in certain derivatives of the metric are permissible.
Consider then the 3 projections of the Einstein tensor ${\cal G}_{AB}$.  
We use the $\epsilon = 1$ cases of the Codazzi and Gauss constraints (\ref{cod}) 
and (\ref{gauss}) for ${\cal G}_{\Gamma \perp}$ and ${\cal G}_{\perp\perp}$ respectively.  
For ${\cal G}_{\Gamma\Delta}$, the following construction is used.

One begins by writing down the contracted Gauss equation (\ref{contG}) 
and subtracts off  $\frac{1}{2}f_{\Gamma\Delta} $ times the doubly-contracted Gauss equation 
--(\ref{Gpp}): 
\be
\begin{array}{c}
{\cal R}_{\Gamma\Delta}                                            \\
- \frac{\cal R}{2}f_{\Gamma\Delta}
\end{array}
\begin{array}{c}
- \\ + 
\end{array}
\begin{array}{c}
{\cal R}_{\perp \Gamma \perp \Delta} = \\
{\cal R}_{\perp\perp}f_{\Gamma\Delta} 
\end{array}
\begin{array}{c}
 R_{\Gamma\Delta}                \\
 -\frac{R}{2}f_{\Gamma\Delta}
\end{array}
\begin{array}{c}
- \\ +
\end{array}
\begin{array}{c}
{\cal K} {\cal K}_{\Gamma\Delta} + { {\cal K}_{\Gamma}} ^{\Lambda}{\cal K}_{\Delta\Lambda} \\
\frac{{\cal K}^2 - {\cal K} \circ {\cal K}}{2}f_{\Gamma\Delta}.
\end{array}
\label{bfbwefes}
\ee 
The following steps are then applied.  

\noindent Step 1: The Ricci equation (\ref{thirdproj}) 
is used to remove all the ${\cal R}_{\perp\Gamma\perp\Delta}$.  

\noindent Step 2: The contracted Ricci equation (\ref{tpcont}) 
is used to remove all the  ${\cal R}_{\perp\perp}$.   Thus   
\be
\left(
\begin{array}{c}
{\cal R}_{\Gamma\Delta}                               \\
- \frac{{\cal R}}{2}f_{\Gamma\Delta}
\end{array}
\right)
\begin{array}{c}
- \\ +
\end{array}
\begin{array}{c}
\left[
\frac{
\delta_{\vec{\beta}} {\cal K}_{\Gamma\Delta} -  D_{\Gamma} D_{\Delta}\alpha
}
{\alpha} + {{\cal K}_{\Gamma}}^{\Pi} {\cal K}_{\Delta\Pi}
\right]                                                         
\\  
\left[
\frac{\delta_{\vec{\beta}} {\cal K} -  D^2\alpha}{\alpha} - {\cal K} \circ {\cal K}
\right] 
f_{\Gamma\Delta}
\end{array}
\mbox{ } \mbox{ } \mbox{ }
 =
\left(
\begin{array}{c}
R_{\Gamma\Delta} \\
-\frac{R}{2}f_{\Gamma\Delta}
\end{array}
\right)
\begin{array}{c}
- \\ +
\end{array}
\begin{array}{c} 
{\cal K} {\cal K}_{\Gamma\Delta} + { {\cal K}_{\Gamma}} ^{\Lambda}{\cal K}_{\Delta\Lambda} 
\\ 
\frac{{\cal K}^2 - {\cal K} \circ {\cal K}}{2}f_{\Gamma\Delta}.  
\end{array}
\label{step0}
\ee
This is then rearranged to form the ``GR CP'' geometrical identity
\be
{\cal G}_{\Gamma\Delta} = G_{\Gamma\Delta} - {\cal K} {\cal K}_{\Gamma\Delta} + 2{ {\cal K}_{\Gamma}} ^{\Lambda}{\cal K}_{\Delta\Lambda}   
                                                                     + \frac{{\cal K}^2 + {\cal K} \circ {\cal K}}{2}f_{\Gamma\Delta}
                                                                     + \frac{\delta_{\vec{\beta}} {\cal K}_{\Gamma\Delta} -  D_{\Gamma} D_{\Delta}\alpha 
                                                                     - (\delta_{\vec{\beta}} {\cal K} -  D^2\alpha)f_{\Gamma\Delta}} {\alpha}.
\label{GRCPid}
\ee
$$\mbox{Then performing} \mbox{ } \begin{array}{c} \mbox{lim} \\ \epsilon \longrightarrow 0\end{array} \int_{-\epsilon}^{+\epsilon}{\cal G}_{AB}dz \mbox{ } 
\mbox{one obtains the j.c's} 
\mbox{ } \mbox{ }\mbox{ } \mbox{ }\mbox{ } \mbox{ }\mbox{ } \mbox{ }\mbox{ } \mbox{ }\mbox{ } \mbox{ }\mbox{ } \mbox{ }\mbox{ } \mbox{ }
\mbox{ } \mbox{ }\mbox{ } \mbox{ }\mbox{ } \mbox{ }\mbox{ } \mbox{ }\mbox{ } \mbox{ }\mbox{ } \mbox{ }\mbox{ } \mbox{ }\mbox{ } \mbox{ }
\mbox{ } \mbox{ }\mbox{ } \mbox{ }\mbox{ } \mbox{ }\mbox{ } \mbox{ }\mbox{ } \mbox{ }\mbox{ } \mbox{ }\mbox{ } \mbox{ }\mbox{ } \mbox{ }
\mbox{ } \mbox{ }\mbox{ } \mbox{ }\mbox{ } \mbox{ }\mbox{ } \mbox{ }\mbox{ } \mbox{ }\mbox{ } \mbox{ }\mbox{ } \mbox{ }\mbox{ } \mbox{ }
\mbox{ } \mbox{ }\mbox{ } \mbox{ }\mbox{ } \mbox{ }\mbox{ } \mbox{ }\mbox{ } \mbox{ }\mbox{ } \mbox{ }\mbox{ } \mbox{ }\mbox{ } \mbox{ } $$
\be
[{\cal G}_{\perp\perp}]_-^+ = 0 \mbox{ } , \mbox{ } \mbox{ } [{\cal G}_{\Gamma\perp}]_-^+ = 0 \mbox{ },
\label{jc0}
\ee
\be
[{\cal G}_{\Gamma\Delta}]_-^+ =  [{\cal K}_{\Gamma\Delta} - f_{\Gamma\Delta}{\cal K}]^+_-.   
\label{primjn}
\ee
The derivation of this last equation makes use of normal coordinates (in which case the hypersurface derivative $\delta_{\vec{\beta}}$ becomes the normal derivative 
$\frac{\pa}{\pa z}$) and the rearrangement to the `Israel' geometrical identity 
\be
{\cal G}_{\Gamma\Delta} = G_{\Gamma\Delta} + {\cal K} {\cal K}_{\Gamma\Delta} + 2{ {\cal K}_{\Gamma}} ^{\Lambda}{\cal K}_{\Delta\Lambda}   
                                                                     + \frac{{\cal K}^2 + {\cal K} \circ {\cal K}}{2}f_{\Gamma\Delta}
                                                                     + \frac{\pa}{\pa z}\left({\cal K}_{\Gamma\Delta} -  f_{\Gamma\Delta}{\cal K} \right) 
\label{Israel}
\ee
via the definition of extrinsic curvature (\ref{ecd}) to form the complete normal derivative 
$\frac{\pa}{\pa z}({\cal K}_{\Gamma\Delta} - {f}_{\Gamma\Delta}{\cal K})$ .

\noindent Step 3: One then further assumes that the (4, 1)-dimensional EFE's  
${\cal G}_{\Gamma\Delta} = {\cal T}_{\Gamma\Delta}$ hold.  If one then additionally 
$$\mbox{uses the thin matter sheet energy-momentum }
{\cal Y}_{AB} = 
\begin{array}{c}
\mbox{lim} \\ 
\epsilon \longrightarrow 0
\end{array}
 \int_{-\epsilon}^{+\epsilon} {\cal T}_{AB} dz,  
\mbox{ one obtains the j.c's } \mbox{ } \mbox{ } \mbox{ } \mbox{ } \mbox{ } \mbox{ } \mbox{ } \mbox{ } \mbox{ } \mbox{ } \mbox{ } \mbox{ } 
                               \mbox{ } \mbox{ } \mbox{ } \mbox{ } \mbox{ } \mbox{ } \mbox{ } \mbox{ } \mbox{ }$$
\be
0 = {\cal Y}_{\perp\perp} \mbox{ } , \mbox{ } \mbox{ } 0 = {\cal Y}_{\Gamma\perp} \mbox{ },
\label{jc0M}
\ee
\be
[{\cal K}_{\Gamma\Delta}]_-^+ =  
\left(
{\cal Y}_{\Gamma\Delta} - \frac{{\cal Y}}{D - 2}f_{\Gamma\Delta}
\right)
\label{prez2}
\ee
(performing a trace-reversal to obtain the last equation).

\subsection{SMS's Braneworld EFE's}

We next recall the method SMS use to obtain their BEFE's \cite{SMS}.    
They begin by forming the (3, 1)-dimensional Einstein tensor $ G_{\Gamma\Delta}$ just like we 
obtain (\ref{bfbwefes}) above.
SMS then apply three steps to this equation.

\noindent Step S1: Using the definition of the Weyl tensor, 
${\cal R}_{\perp \Gamma \perp \Delta }$  is replaced by the electric part of the Weyl tensor, 
${\cal E}_{\Gamma\Delta} \equiv {\cal C}_{\perp \Gamma \perp \Delta} $ and extra terms built from the projections of ${\cal R}_{AB}$.  

\noindent Step S2 ( = Step 3 of the above subsection): 
The (4, 1)-dimensional EFE's are then assumed, which permits one to exchange all remaining projections of ${\cal R}_{AB}$ for (4, 1)-dimensional 
energy-momentum terms.  Only when this is carried out does (\ref{bfbwefes}) become a system of field equations rather than of geometrical identities.    
We refer to the field equations at this stage as ``timelike hypersurface EFE's'' (THEFE's),\fn{This choice of name reflects their 
superficial resemblance to the (3, 1)-dimensional EFE's, although as discussed below, this resemblance is often only superficial, including in the SMS case.} as opposed to the 
braneworld EFE's which arise at the next stage.  

\noindent Step S3: A special subcase of THEFE's are braneworld EFE's (BEFE's), which are obtained in normal coordinates by choosing the (thin) braneworld energy-momentum 
tensor ansatz 

\be
{\cal T}_{AB} = {\cal Y}_{AB}\delta(z) - \Lambda g_{AB} \mbox{ } , \mbox{ } \mbox{ } 
{\cal Y}_{AB} \equiv (T_{AB} - \lambda f_{AB} ) \mbox{ } , \mbox{ } \mbox{ } T_{AB}z^A = 0, 
\label{BWEM}
\ee
where $T_{\Gamma\Delta}$ is the energy-momentum of the matter confined to the brane. This is a specialization due to the specific presence of 
(3, 1) and (4, 1) cosmological constants $\lambda$ and $\Lambda$, and by $\Lambda$ being the only bulk contribution.  A more precise formulation of this, 
and generalizations, are the subject of Sec 4.5.

One then adopts the j.c's (\ref{jcf}), (\ref{jc0}), and (\ref{prez2}) with the additional supposition of $Z_2$ symmetry\fn{The difference 
in the sign of (\ref{prez2}) between this paper and SMS's paper is due 
to our use of the opposite sign convention in the definition of extrinsic curvature.
We compensate for this in subsequent formulae by also defining $K_{ab} = -K_{ab}^+$ rather than $+K_{ab}^+$.}:  
\be
- {\cal K}_{\Gamma\Delta} \equiv {\cal K}_{\Gamma\Delta}^+ = - {\cal K}_{\Gamma\Delta}^- \mbox{ } \mbox{ } \Rightarrow \mbox{ } \mbox{ }
{\cal K}_{\Gamma\Delta} = - \frac{\kappa_5^2}{2}\left({\cal Y}_{\Gamma\Delta} - \frac{{\cal Y}}{3}f_{\Gamma\Delta}\right) = 
- \frac{\kappa_5^2}{2}\left(T_{\Gamma\Delta} - \frac{{T - \lambda}}{3}f_{\Gamma\Delta}\right), 
\label{41jc}
\ee
where the 5-dimensional gravitational constant $\kappa_5^2$ has been made explicit.
Then SMS's BEFE's read
\be
G_{\Gamma\Delta} = L^{\mbox{\scriptsize SMS\normalsize}}_{\Gamma\Delta} + Q^{\mbox{\scriptsize SMS\normalsize}}_{\Gamma\Delta} - {\cal E}_{\Gamma\Delta}, 
\ee
where $Q_{\Gamma\Delta}(T)$ and $L_{\Gamma\Delta}(T)$ are the terms quadratic in, and linear together with zeroth order in $T$
respectively, given by 
\be
Q^{\mbox{\scriptsize SMS\normalsize}}_{\Gamma\Delta} = \kappa_5^4
\left[ 
\frac{T}{12}T_{\Gamma\Delta} - \frac{1}{4}T_{\Gamma\Pi}{T^{\Pi}}_{\Delta} + 
\left(
\frac{T \circ T}{8} - \frac{T^2}{24}
\right)
f_{\Gamma\Delta}
\right] 
\label{BEFE}
\ee
\be
L_{\Gamma\Delta}^{\mbox{\scriptsize SMS\normalsize}} = -\frac{\kappa_5^2}{2}\left(\Lambda + \frac{\kappa_5^2}{6}\lambda^2\right)f_{\Gamma\Delta} + 
\frac{\kappa^4_5}{6}\lambda T_{\Gamma\Delta}.
\ee
As opposed to the (3, 1)-dimensional EFE's, SMS's BEFE's are not 
closed since they contain the unspecified electric part of the Weyl tensor 
${\cal E}_{\Gamma\Delta}$. Although it also contains 15 equations, the SMS BEFE--Gauss--Codazzi 
system\fn{The 5 Gauss--Codazzi equations may be seen as consistency conditions on the
brane. For $j_{\Gamma} = 0$ the 4 Codazzi equations correspond to energy-momentum conservation 
on the brane.} is not equivalent to the (4, 1)-dimensional EFE's: indeed SMS write down further 
third-order equations for the ``evolution'' away from the timelike brane of ${\cal E}_{\Gamma\Delta}$, 
by use of the $z$-derivative of the contracted Gauss equation (\ref{contG}), Bianchi identities and the Ricci equation (\ref{thirdproj}).  
This then involves the magnetic part\fn{As the 5-dimensional alternating tensor has 5 indices, one has 
two choices for the number of indices in what is to be taken to be the definition of 
the magnetic Weyl tensor.  The above is the 3-indexed definition; the other possible definition 
is $H^{\Lambda\Delta} = \epsilon^{\Lambda\Delta\Gamma\Sigma\Pi}B_{\Gamma\Sigma\Pi}$.} of the Weyl tensor 
${\cal B}_{\Lambda\Delta\Gamma} \equiv {\cal C}_{\perp \Gamma\Delta\Lambda}$, 
the ``evolution" of which follows from further Bianchi identities.    
This full brane-bulk SMS system is then closed. 

\noindent Step 4: In practice, however, instead of the difficult treatment of
this third-order system, other practitioners have often worked on the SMS 
BEFE's alone.  
This involves either the ad hoc prescription of the functional form of  
${\cal E}_{\Gamma\Delta}$ (sometimes taken to be zero\fn{It is zero for example 
when one makes the restrictive presupposition of
a conformally-flat bulk, such as in AdS spacetime.} ).  
In fact it is often first decomposed according to a standard procedure 
\cite{MWBH, Maartensdec}.  
Because the original functional form is
unknown, the functional forms of each of the parts
defined by the decomposition is also unknown.
Some of these parts are set equal to zero whereas other parts are taken to
have other functional forms 
\fn{For example the `Weyl charge' for black holes in \cite{BH}.} 
(in particular a radiation fluid term).    
These terms are then argued to be small in the circumstances arising in the 
inflationary \cite{MWBH} and perturbative \cite{Maartensdec} treatments.
However other interpretations may be possible.  
Having dealt with
${\cal E}_{\Gamma\Delta}$ in one of the above ways, the form (53) of
$Q_{\Gamma\Delta}^{\mbox{\scriptsize SMS\normalsize}}$ is then often taken to
be uniquely defined and the starting-point of many works on brane cosmology
\cite{Langlois, otherpapers2, otherpapers1, Coley1}.

However, SMS's procedure is far from unique.  It turns out that there are many reformulations of the BEFE arising from geometrical identities.  
Each has a distinct split of the non-Einsteinian BEFE terms into `bulk' and `brane' terms.  Whereas all these formulations are clearly equivalent, their 
use helps clarify how to interpret SMS's braneworld.  Were one to truncate the `bulk' terms in each case (in direct analogy with the usual practice of throwing away 
the Weyl term), then the BEFE's obtained in each case would generally be inequivalent.

\subsection{Ambiguity in the Formulation of the BEFE's}

The Weyl term in SMS's BEFE's has been the subject of much mystery.  
How should it be interpreted?  
Is it right to throw it away and if so under which circumstances?      
We emphasize that our stance is broader than merely about what functional form is allocated to 
the Weyl term (e.g whether it is zero everywhere).  It is about how formulations can be chosen 
in which the BEFE's are not explicitly formulated in terms of a Weyl term.  We first remove 
some misconceptions as to how SMS's procedure leads to a BEFE containing a Weyl term.  
Does it have anything to do with the modelling of braneworld scenarios?  
No, for the Weyl term is already in the SMS THEFE before the braneworld energy-momentum ansatz 
is invoked.  Furthermore,  all the procedures used in SMS's method are independent of signature 
and dimension.  Thus this issue of a Weyl term must have already arisen long ago in the study 
of the GR CP.  
So why is there no manifest Weyl term piece in the GR CP formulation of the EFE's?  
The answer is simple.  Indeed, we have already seen the answer in Sec 4.1 since it 
is fundamentally tied to how the crucial junction condition (\ref{prez2}) is obtained in the 
first place!. In the ``GR CP'' and Israel procedures, \it one uses the Ricci equation 
(\ref{thirdproj}) to remove the ${\cal R}_{\perp \Gamma \perp \Delta}$ term. \normalfont 
If there is no early use of the Ricci equation, one is left with ${\cal E}_{\Gamma\Delta}$ in 
the THEFE's, which requires later use of the Ricci equation to ``evolve'' it.  

So there is a choice as to whether one formulates the BEFE's with or without an explicit Weyl 
term.  In the usual treatment of the split of the EFE's (Sec 2.2), one does not use an 
explicit Weyl term.  In the (4, 1)-d case, this gives a well-understood system of 15 p.d.e's 
in the variables $f_{\Gamma\Delta}$.  The option of using an explicit Weyl term gives a 
considerably larger, more complicated system of p.d.e's with variables $f_{\Gamma\Delta}$, 
${\cal E}_{\Gamma\Delta}$ and ${\cal B}_{\Gamma\Delta\Lambda}$. 

In fact a similar scheme known as the \it threading formulation \normalfont \cite{Ellis}
is sometimes used in the usual (3, 0; --1) application.  The idea behind this formulation is to 
treat as primary the geodesic congruences perpendicular to the foliation 
rather than the foliation itself.  One then uses only that information on the hypersurfaces 
that arrives along the impinging geodesic congruence, on the grounds that this is the 
physically-significant information.  Thus it is a `deliberately incomplete system' from the 
foliation perspective.  
Whereas this and the SMS formulation are similar in their use of Weyl variables, 
the SMS system does not appear to be a threading formulation. 
Also, in any case the idea 
of having a deliberately partial system in the threading formulation is clearly tied to 
signature-specific physical reasons which do not carry over to the signature relevant to SMS's 
equations.  We also note that some other third-order reformulations of the $(3, 0; -1)$ split 
of the EFE's are sometimes used to seek to cast the EFE's into hyperbolic forms that manifestly have 
theorems associated with them \cite{RF}.  Whereas this is 
precisely the sort of result that is spoiled by considering instead a sideways split\fn{Having 
argued that the second-order sideways ``GR CP'' is not known to be well-posed, we should add 
that it is unlikely that third-order formulations are likely to be better-behaved in this 
respect.}, it serves to illustrate that what at first seems a `mere reformulation' of a set 
of equations can in fact be used to prove highly nontrivial theorems.  
So, whereas similar complicated formulations have been used elsewhere in the GR 
literature, SMS's unstated motivation to have a complicated formulation does not coincide with 
the motivation elsewhere in the literature.  Below we bring attention to many reformulations of 
SMS's system, so we ask: what is the motivation for the original SMS formulation?  
Should the use of some simpler second-order formulation be preferred?  
Is SMS's formulation or any other third- or second-order formulation singled out by good behaviour, 
either in general or for some particular application?   

Also, from first principles the SMS procedure to obtain their BEFE is  
quite complicated. For, since they use the j.c obtained by the Israel procedure,  
their procedure actually entails beginning with the whole Israel procedure (Steps 1 to 3 of 
Sec 4.1), and then choosing to reintroduce ${\cal R}_{\Gamma\perp\Delta\perp}$ and 
${\cal R}_{\perp\perp}$ by reverse application of Steps 1 and 2.  This is followed by the Weyl 
rearrangement (Step S1), the use of the EFE's (Step S2) and the  substitution of the j.c  
into the extrinsic curvature terms in the braneworld ansatz (Step S3).  
However, despite being complicated, all is well with SMS's scheme since 
any BEFE's obtained by other such combinations of careful procedures will always be equivalent 
because the different steps are related by geometrical identities.  

Step 4 however is not an instance of being careful as it is a truncation.    
Our first point is that whereas in SMS's formulation the non-Einsteinian terms in the BEFE 
might be regarded as a bulk-like ${\cal E}_{\Gamma\Delta}$ and a term quadratic in the brane 
energy-momentum, in other formulations the content of these two terms can be mixed up.  
In general, BEFE's contain a group of non-Einsteinian `bulk' terms we denote by $B_{\Gamma\Delta}$ 
(which include both Weyl terms and  
normal derivatives of objects such as the extrinsic curvature), and a group of non-Einsteinian 
`brane' terms that depend on the brane energy-momentum.  
Thus any temptation to discard the Weyl term in the SMS formulation 
(on the grounds that it involves the unknown bulk over which one has no control) 
should be seen in the light that 
if one considered instead a reformulation, 
then there would be a similar temptation to discard the corresponding 
`bulk' term, which would generally lead to something \it other \normalfont than the Weyl term being discarded.  
Thus for each formulation, the corresponding truncation of the `bulk term' would result in 
inequivalent residual `braneworld physics'.  This is because there are geometrical identities 
that relate `bulk' and `brane' terms, so that the splits mentioned above are highly non-unique 
and thus not true splits at all.  We take this as a clear indication that any such truncations
should be avoided in general.  
Instead, the full system must be studied.

Our second point is that each possible bulk spacetime may contain some hypersurface on  
which a given $B_{\Gamma\Delta}$ vanishes.  Then if one identifies this hypersurface with the 
position of the brane, one has a solution of the full brane-bulk system and not a truncation.  
For example, in any conformally-flat bulk, by definition ${\cal C}_{ABCD} = 0$ and therefore 
$B^{\mbox{\scriptsize SMS\normalsize}}_{\Gamma\Delta} = -{\cal E}_{\Gamma\Delta} = 0$ 
on all hypersurfaces.  Thus any of these could be identified as 
a brane to form a genuine (rather than truncated) $-{\cal E}_{\Gamma\Delta} = 0$ braneworld.  
From this, we can see that \it the SMS formulation is particularly well adapted for the study of 
conformally-flat bulks such as pure AdS\normalfont.  This motivates SMS's formulation as regards this 
common application.  However, also consider repeating the above procedure   
with some $B_{\Gamma\Delta} \neq - {\cal E}_{\Gamma\Delta}$.  This would correspond to a 
genuine (rather than truncated) braneworld model with distinct braneworld physics from that given 
by SMS's particular quadratic term.  
Note that given a model with some $B_{\Gamma\Delta} = 0$,
the BEFE formulation for which $B_{\Gamma\Delta}$ is the bulk term is particularly well adapted for the study of that model.  
Thus different formulations may facilitate the study of braneworlds with different 
braneworld physics.  In the context of conformally-flat spacetimes, it is probably true that the 
$-{\cal E}_{\Gamma\Delta} = 0$ braneworlds outnumber the braneworlds for which any other (or 
even all other) $B_{\Gamma\Delta} = 0$ since these other conditions 
are not automatically satisfied on all embedded hypersurfaces.  
Rather, each of these other conditions constitutes a difficult geometrical problem, 
somewhat reminiscent of the question of which spacetimes contain a maximal (K = 0) or CMC slice 
\cite{MT}.  However, generic spacetimes are not conformally-flat.  
For a generic spacetime, we see no difference between the status of the condition 
`$-{\cal E}_{\Gamma\Delta} = 0$ on some hypersurface' and the condition `any other particular 
$B_{\Gamma\Delta} = 0$ on some hypersurface'.  Because braneworlds constructed in 
each of these cases have a different residual quadratic term and thus a propensity to have distinct  
braneworld physics, and because we do not know how frequently each of these cases occur,  
we question whether anything inferred from conformally-flat models with the SMS quadratic term 
need be typical of the full SMS brane-bulk system.  
Confirmation of this would require study of the full range of difficult geometrical problems 
$B_{\Gamma\Delta} = 0$, and the construction of concrete examples of 
non-SMS quadratic term braneworld models together with the assessment of whether their 
braneworld physics is conceptually and observationally acceptable.  
  
For the moment we study what is the available range of reformulations and thus of 
$B_{\Gamma\Delta}$.  To convince the reader that such reformulations exist,  
we provide a first example before listing all the steps which are available for 
reformulating the BEFE's.      

Assume we do not perform all the steps implicit within SMS's work but rather just the Israel 
steps to obtain the j.c and then use it in the field equation (\ref{Israel}) that gave rise to 
it (as done in \cite{ATlett}), or (as done below) use it in the ``GR CP'' field equation 
following from (\ref{GRCPid}).  
In other words, why not apply the braneworld ansatz to e.g the Israel or ``GR CP'' formulations 
rather than to the SMS formulation?    In the ``GR CP'' case we then obtain
\be
G_{\Gamma\Delta} = L_{\Gamma\Delta} + Q_{\Gamma\Delta} + B_{\Gamma\Delta} \mbox{ , with }
\label{CPBEFE}
\ee 
\be
Q_{\Gamma\Delta} = -\frac{\kappa_5^4}{72}
\left[
36{T_{\Gamma}}^{\Pi}T_{\Pi\Delta} - 18TT_{\Gamma\Delta} + (9T\circ T + T^2)f_{\Gamma\Delta}
\right]
\label{QCP}
\ee
\be
L_{\Gamma\Delta} = \frac{\kappa_5^4}{9}(T \lambda  - 2\lambda^2) f_{\Gamma\Delta}   
+ \kappa_5^2[T_{\Gamma\Delta} - (\lambda + \Lambda) f_{\Gamma\Delta}]
\label{LCP}
\ee
\be
B_{\Gamma\Delta} = f_{\Gamma\Delta}\frac{\pa K}{\pa z} - \frac{\pa K_{\Gamma\Delta}}{\pa z}.  
\label{BCP}
\ee
This example serves to illustrate that choosing to use a different formulation can cause the 
`brane' quadratic term $Q_{\Gamma\Delta}(T)$ to be different.  Also note that this formulation 
makes no explicit use of the Weyl term.  Thus this BEFE, along with the Gauss and Codazzi 
constraints, forms a small second-order system, in contrast with the much larger third-order 
SMS system.   

Now we further study the list of steps \cite{ATlett} which may be applied in the construction 
of BEFE's.  

\noindent Steps S1 and 3 together mean that the Weyl 
`bulk' term ${\cal E}_{\Gamma\Delta}$ is equivalent to 
the Riemann `bulk' term together with matter terms.  
This swap by itself involves no terms which are 
quadratic in the extrinsic curvature.

\noindent Step 1 says that the Riemann `bulk' term is equivalent 
to the hypersurface derivative of the extrinsic 
curvature together with a ${\cal K}_{\Gamma\Pi}{{\cal K}^{\Pi}}_{\Delta}$ term.

\noindent Steps S1 and 3 together say that the hypersurface derivative  
of the trace of the extrinsic curvature is equivalent to a matter term together with 
a ${\cal K} \circ {\cal K}$  term.  

Furthermore, one can use both Steps 2-3 and Step 1, on
arbitrary proportions (parametrized by $\mu$ and $\nu$)
of ${\cal R}_{\perp \Gamma\perp \Delta}$ and of ${\cal R}_{\perp\perp}$:

\noindent
\be
\begin{array}{ll}
G_{\Gamma\Delta} & = {\cal G}_{\Gamma\Delta} - (1 +\nu){\cal R}_{\perp \Gamma\perp \Delta}\mbox + (1 - \mu){\cal R}_{\perp\perp}f_{\Gamma\Delta} 
+\frac{1}{\alpha}
\left[
\nu(\delta_{\vec{\beta}} {\cal K}_{\Gamma\Delta} -  D_{\Gamma} D_{\Delta}\alpha)  
+ \mu(\delta_{\vec{\beta}}  {\cal K} -  D^2\alpha)f_{\Gamma\Delta}
\right] \\&
+ 
{\cal K} {\cal K}_{\Gamma\Delta} + (\nu - 1) {{\cal K}_{\Gamma}}^{\Pi} {\cal K}_{\Delta\Pi} 
- \frac{{\cal K}^2}{2}f_{\Gamma\Delta}  +  (\frac{1}{2} - \mu) {\cal K} \circ {\cal K}f_{\Gamma\Delta}
\end{array} 
\label{2param}
\ee
This introduces freedom in the coefficients of the
${\cal K}_{\Gamma\Pi}{{\cal K}^{\Pi}}_{\Delta}$ and ${\cal K} \circ {\cal K}$ contributions to 
the quadratic term $Q_{\Gamma\Delta}({\cal K})$ in the THEFE's.  
We next find further freedom in $Q_{\Gamma\Delta}$ by choice of the objects to be regarded as primary.  

We are free to choose a `bulk' term described by hypersurface derivatives $\delta_{\vec{\beta}} $ (which are partial derivatives $\frac{\pa}{\pa z}$ 
in normal coordinates) of objects related to the extrinsic curvature ${\cal K}_{\Gamma\Delta}$ 
by use of the metric tensor (including its inverse and determinant $f$).  
The underlying reason for doing this is that it is just as natural to treat such an object, 
rather than the extrinsic curvature itself, as primary (see below for examples).  

Upon careful consideration, there are three separate ways such objects can be related 
to the extrinsic curvature: raising indices, removing a portion of the trace by defining 
$_{\eta}{\cal K}_{\Gamma\Delta} \equiv {\cal K}_{\Gamma\Delta} - \eta{\cal K}f_{\Gamma\Delta}$, 
and densitizing by defining $^{\xi}{\cal K}_{\Gamma\Delta} \equiv f^{\xi} \times {\cal K}_{\Gamma\Delta}$.  
The hypersurface derivatives of these objects are related to those of the extrinsic curvature by
\be
\delta_{\vec{\beta}} {\cal K}_{\Gamma\Delta} = \delta_{\vec{\beta}}  (f_{\Gamma\Pi}{K^{\Pi}}_{\Delta}) = f_{\Gamma\Pi}\delta_{\vec{\beta}}  {{\cal K}^{\Pi}}_{\Delta} - 2\alpha {\cal K}_{\Gamma\Pi}{{\cal K}^{\Pi}}_{\Delta},
\label{i}
\ee
\be
\delta_{\vec{\beta}}{\cal K}_{\Gamma\Delta} = \delta_{\vec{\beta}}{\cal K}^{\mbox{\scriptsize $\eta$\normalsize}}_{\Gamma\Delta} + \eta
\left(\delta_{\vec{\beta}}{\cal K}f_{\Gamma\Delta} - 2\alpha{\cal K}{\cal K}_{\Gamma\Delta}
\right),
\label{64}
\ee    
\be
\delta_{\vec{\beta}}{\cal K}_{\Gamma\Delta} = (\mbox{}f)^{-\xi_1}\delta_{\vec{\beta}}(\mbox{}f^{\xi_1}{\cal K}_{\Gamma\Delta}) + 2\alpha\xi_1{\cal K}{\cal K}_{\Gamma\Delta}.
\label{65}
\ee  
Further useful equations arise from the traces of these:  
\be 
\delta_{\vec{\beta}} {\cal K} = f^{\Gamma\Delta} \delta_{\vec{\beta}}{\cal K}_{\Gamma\Delta} + 2\alpha {\cal K} \circ {\cal K},
\label{63}
\ee
\be
\delta_{\vec{\beta}}{\cal K} = \frac{1}{1 - 4\eta}[f^{\Gamma\Delta}\delta_{\vec{\beta}}\mbox{}_\eta{\cal K}_{\Gamma\Delta} - 2\alpha\eta({\cal K}^2 - {\cal K}\circ{\cal K})] 
\mbox{ } , \mbox{ } \mbox{ } \eta \neq \frac{1}{4}
\label{new}
\ee
\be
\delta_{\vec{\beta}}{\cal K} = (\mbox{}f)^{-\xi_2}\delta_{\vec{\beta}}(\mbox{}f^{\xi_2}{\cal K}) + 2\alpha\xi_2{\cal K}^2, 
\label{66}
\ee
where the $\delta_{\vec{\beta}}{\cal K}$ in (\ref{new}) and (\ref{new}) has been obtained via (\ref{66}).  

The following examples of $_{\eta}^{\xi}{\cal K}_{\Gamma\Delta}$ illustrate that the use of 
such objects is entirely natural:  $_{{1/4}}^{0}{\cal K}_{\Gamma\Delta}$ 
is the ${\cal K}^{\mbox{\scriptsize T\normalsize}}_{\Gamma\Delta}$ commonly used in the IVP literature,  
and $_{0}^{{1/2}}{\cal K}_{\Gamma\Delta}$ appears in the guise of forming the complete normal derivative in the Israel procedure.  
Also, the ``gravitational momenta'' are $p_{\Gamma\Delta} \equiv  - _{1}^{{1/2}}{\cal K}_{\Gamma\Delta}$.

The above thorough consideration of possible `bulk' terms permits all 
four THEFE terms homogeneously quadratic in the extrinsic curvature to be changed 
independently.  One may think that we have a redundancy in providing 8 ways to change only 4 
coefficients.  
However, one can afford then to lose some of the freedoms by making extra 
demands, of which we now provide four examples of relevance to this paper.  
First, one could further demand that there is no Weyl term in the THEFE's (as discussed in Sec. 4.4).  
Second, unequal densitization of ${\cal K}_{\Gamma\Delta}$ and ${\cal K}$ ($\xi_1 \neq \xi_2$) 
corresponds to interpreting the fundamental variable to be some densitization of the metric 
rather than the metric itself.  Whereas this is again a common practice (for example the scale-free metric  
of the IVP literature is ${   f^{  -\frac{1}{n} }   }{   f_{\Gamma\Delta}   }$ in dimension $n$), 
the use of such an object as fundamental variable does appear to complicate the isolation of 
the (3, 1) Einstein tensor truly corresponding to this fundamental variable.  Thus this option is not pursued in this paper.  
Third, one may start by declaring that one is to use particular well-known primary objects (such as the 
``gravitational momenta'') and still desire to be left with much freedom of formulation.    
Fourth, one could declare that one is to use the raised objects given by (\ref{63}), in which case 
the further ability to change coefficients by use of (\ref{64}) is lost, 
since moving a Kronecker delta rather than a metric through the derivative clearly generates no terms quadratic in 
the extrinsic curvature.

\subsection{Examples of Formulations of BEFE's with no Quadratic Terms}

As a consequence of the above freedoms, there are many formulations in which all four
coefficients vanish, and hence $Q_{\Gamma\Delta}({\cal K}) = 0$.       
From this it follows that $Q_{\Gamma\Delta}(T)$ is zero [and it is easy to show that all 
instances of $Q_{\Gamma\Delta}(T) = 0$ follow from $Q_{\Gamma\Delta}({\cal K}) = 0$].  
Thus it suffices to seek for cases of THEFE's with $Q_{\Gamma\Delta}({\cal K}) = 0$ 
to obtain all cases of BEFE formulations that have no quadratic term.  
We now motivate these formulations and then choose to exhibit three that comply with 
some of the extra demands in the previous paragraph.

The diversity of `brane'-`bulk' splits ensures that truncations such as Step 4 
produce all possible combinations of quadratic terms as residual `braneworld physics'.  
Alternatively, one may suspect that there might be solutions to the full brane-bulk system 
that just happen to have a particular $B_{\Gamma\Delta} = 0$ on some hypersurface 
which is then identified as a brane.   We speculate that each of these situations will often 
lead to different answers to questions of physical interest.  Whereas most 
Friedmann--Lema\^{i}tre--Robertson--Walker (FLRW) perfect fluid 
models with 

\noindent equation of state $P = (\gamma - 1)\rho$ arising thus will be similar, 
differences will be more salient in models with more complicated equations of state, 
in perturbations about FLRW  (as started in \cite{Maartensdec, SLMW, MB} in the SMS-adapted 
case) and in anisotropic models (as started by \cite{Coley1, Coleyconj} in the SMS-adapted case).  
These in turn constitute natural frameworks to seriously justify the late-time emergence of FLRW behaviour 
and the likelihood of inflation \cite{otherpapers1} as well as the study of singularities 
\cite{Coleyconj, MB} on the brane.  
We emphasize that, 
for a satisfactory study of whether any particular full brane-bulk (as opposed to truncated) case 
leads to any differences from the hitherto-studied $-{\cal E}_{\Gamma\Delta} = 0$ case, 
one would require a full brane-bulk solution explicitly constructed to satisfy some $B_{\Gamma\Delta} = 0$ 
on some hypersurface within a particular given bulk spacetime.  
Since we currently have no such example, our arguments currently only 
support the far simpler idea that truncation should be avoided.    

We illustrate that in different formulations, the truncation of the corresponding `bulk' terms 
can lead to 
big differences in the residual `braneworld physics', without any of the above 
lengthy calculations. We do this by formulating the `bulk' part so that there is no 
corresponding $Q$ term at all.  Thus these truncations give  the `$\rho$' of standard FLRW 
cosmology rather than the `$\rho$ and $\rho^2$' of brane cosmology  
\cite{Langlois, MWBH, Maartensdec}.  As a result whether we have a `$\rho$ and $\rho^2$' 
brane cosmology depends on the choice of formulation.    So we argue that since the SMS 
procedure followed by truncating the Weyl term is a hitherto unaccounted-for choice out of 
many possible procedures, then adopting the particular homogeneous quadratic term of SMS (often taken as the 
starting-point of brane cosmology) appears to be unjustified.  Rather, we conclude that no 
particular truncation should be privileged as the act of truncation imprints undesirable 
arbitrariness into the study of the truncated system.  Whereas the (3, 1)-dimensional trace 
of ${\cal E}_{\Gamma\Delta}$ happens to be zero (including use of  antisymmetry) and thus 
might\fn{This is if one treats the geometric content of the Weyl tensor as an effective or 
`induced' energy--momentum, and furthermore assumes that this ought to behave like a perfect 
fluid.  Maartens' decomposition of ${\cal E}_{\Gamma\Delta}$ \cite{Maartensdec}, 
based on the first assumption, permits the removal of the second assumption.} 
phenomenologically look like pure radiation fluid to observers on the brane, 
other bulk characterizations would typically not look like a pure radiative fluid.  
This may open up phenomenological possibilities.  

Also, before further study of SMS's full (untruncated, third-order)
system is undertaken, some of the reformulations along the lines
suggested in Sec 4.4 might turn out to be more tractable.
In particular those reformulations which fully eliminate
${\cal R}_{\perp \Gamma\perp \Delta}$ by the early use of the Ricci equation are already
closed as second-order systems. These include the Israel formulation in \cite{ATlett},
the ``GR CP'' formulation (\ref{CPBEFE}--\ref{BCP}), and our second and
third examples below, which contain neither a Weyl term nor a quadratic term.

For our first example, we take as the primary object the antidensitized extrinsic curvature
$\underline{K}_{ab} \equiv \frac{   K_{ab}   }{   \sqrt{ h  }   }$
so that the `bulk' term is (partly) a combination of this object's normal derivatives. 
The corresponding BEFE's are: 
\be
G_{\Gamma\Delta} = {\cal L}_{\Gamma\Delta} + B_{\Gamma\Delta}
\ee
\be
L_{\Gamma\Delta} = \frac{{\cal T}_{\Gamma\Delta}}{3} + \frac{1}{6}\left(5{\cal T}_{\perp\perp} - 
{\cal T}\right)f_{\Gamma\Delta} \mbox{ } , \mbox{ } \mbox{ }
B_{\Gamma\Delta} = -2{\cal E}_{\Gamma\Delta} + \sqrt{\mbox{}f}
\left(
\frac{\pa {\underline{\cal K}_{\Gamma\Delta}}}{\pa z}
- \frac{1}{2}f^{\Gamma\Delta}
\frac{\pa \underline{\cal K}_{\Gamma\Delta}}{\pa z} f_{\Gamma\Delta}
\right)
\ee
(where we have chosen to remove all projections of ${\cal R}_{AB}$ by the EFE's).

To derive this, take (\ref{2param}) in normal coordinates.  
Choose to convert all of the $\frac{\pa {\cal K}}{\pa z}$ into $f^{\Lambda\Sigma}\frac{\pa {\cal K}_{\Lambda\Sigma}}{\pa z}$ by (\ref{63}):
\be
\begin{array}{ll}
G_{\Gamma\Delta} & = {\cal G}_{\Gamma\Delta} - (1 +\nu){\cal R}_{\perp \Gamma\perp \Delta} + (1 - \mu){\cal R}_{\perp\perp}f_{\Gamma\Delta} 
+ \left(
\nu \frac{\pa {\cal K}_{\Gamma\Delta}}{\pa z}  
+ \mu\frac{\pa {\cal K}}{\pa z}f_{\Gamma\Delta}
\right) \\&
+ 
{\cal K} {\cal K}_{\Gamma\Delta} + (\nu - 1) {{\cal K}_{\Gamma}}^{\Pi} {\cal K}_{\Delta\Pi} 
- \frac{{\cal K}^2}{2}f_{\Gamma\Delta}  +  (\frac{1}{2} + \mu) {\cal K} \circ {\cal K}f_{\Gamma\Delta}
\end{array} 
\ee
Now choosing the primary object to be some densitized $^{\xi}K_{\Gamma\Delta}$ by (\ref{65}) we have 
\be
\begin{array}{ll}
G_{\Gamma\Delta} = &  {\cal G}_{\Gamma\Delta} - (1 + \nu){\cal R}_{\perp \Gamma\perp \Delta} + (1 - \mu){\cal R}_{\perp\perp} f_{\Gamma\Delta}
+\frac{1}{f^{\xi}}
\left(
\nu \frac{\pa \mbox{}^{\xi}K_{\Gamma\Delta})}{\pa z}  
+ \mu f^{\Lambda\Sigma}\frac{\pa \mbox{}^{\xi}K_{\Lambda\Sigma})}{\pa z}  
\right) \\&
+ 
(1 + 2\nu\xi){\cal K}{\cal K}_{\Gamma\Delta} + (\nu - 1) {{\cal K}_{\Gamma}}^{\Pi} {\cal K}_{\Delta\Pi} +(2\mu\xi - \frac{1}{2}){\cal K}^2f_{\Gamma\Delta}  
+  (\frac{1}{2} + \mu) {\cal K} \circ {\cal K}f_{\Gamma\Delta},
\end{array} 
\ee
so clearly $\mu = -\frac{1}{2}$, $\nu = 1$ and the antidensitization choice of 
weighting $\xi = -\frac{1}{2}$ ensure that $Q_{\Gamma\Delta}({\cal K}) = 0$   

The following examples arose from asking if it is possible to find examples in which neither 
$Q_{\Gamma\Delta}(T)$ nor ${\cal E}_{\Gamma\Delta}$ feature.  We found the following BEFE's: 
\be
G_{\Gamma\Delta} = L_{\Gamma\Delta} + B_{\Gamma\Delta}
\ee
\be
L_{\Gamma\Delta} = {\cal T}_{\Gamma\Delta} + \frac{1}{2}\left({\cal T}_{\perp\perp}- \frac{\cal T}{3}\right)f_{\Gamma\Delta} 
\mbox{ } , \mbox{ } \mbox{ }
B_{\Gamma\Delta} = \frac{1}{\sqrt f}\left( \frac{1}{2}\frac{\pa \overline{\cal K}}{\pa z} f_{\Gamma\Delta} 
- \frac{\pa {\overline {\cal K}}^{\Pi}\mbox{}_{\Gamma}}{\pa z} f_{\Pi\Delta} \right)
\ee
by considering as our primary object the densitized extrinsic curvature with one index raised, 
${\overline {\cal K}}^a\mbox{}_b \equiv \sqrt{\mbox{}h}{{\cal K}^a}_b$.  

We obtained these BEFE's by arguing as follows.  In order for the BEFE's to contain no Weyl 
term, $\nu$ is fixed 

\noindent to be $-1$.  Then the only control over 
${\cal K}_{\Gamma\Pi}{{\cal K}^{\Pi}}_{\Delta}$ is from raising by (\ref{i}).  
It is easy to show that this raising must be applied to the whole 
$\frac{\pa{\cal K}_{\Gamma\Delta}}{\pa z}$ in order for the coefficient of 
${\cal K}_{\Gamma\Pi}{{\cal K}^{\Pi}}_{\Delta}$ to be zero.  
Then using $_{\eta}{{\cal K}^{\Lambda}}_{\Delta}$ does not change any terms quadratic in the 
extrinsic curvature.  Also, use of distinct densities for ${\cal K}$ and 
${{\cal K}^{\Lambda}}_{\Delta}$ does not appear to make sense since both quantities are related 
to ${\cal K}_{\Gamma\Delta}$ by a single use of the inverse metric.  
Although all these restrictions make the outcome unlikely, the use of (\ref{65}) and (\ref{66}) 
alone suffices to obtain the above example:
\be
\begin{array}{ll}
G_{\Gamma\Delta} & = {\cal G}_{\Gamma\Delta} + (1 - \mu){\cal R}_{\perp\perp}f_{\Gamma\Delta} 
+ f^{-\xi}
\left(
- \frac{\pa\mbox{}^{\xi}{\cal K}_{\Gamma\Delta}}{\pa z}  
+ \mu\frac{\pa {\cal K}}{\pa z}f_{\Gamma\Delta}
\right) \\&
+ (1 -2\xi)                              {\cal K} {\cal K}_{\Gamma\Delta} 
+ (2\xi\mu - \frac{1}{2})                {\cal K}^2f_{\Gamma\Delta}  
+ (\frac{1}{2} - \mu)                    {\cal K} \circ {\cal K}f_{\Gamma\Delta}
\end{array} 
\ee
which has no quadratic terms if $\mu = 1/2$ and $\xi = 1/2$ (`densitization' weight).

Another possibility is  to replace $^{\xi}{{\cal K}^{\Lambda}}_{\Delta}$ by 
$_{\eta}^{\xi}{{\cal K}^{\Lambda}}_{\Delta}$.  Although this does not immediately 
do anything about the quadratic terms, if we also convert a portion 
$\pi$ of $\frac{\pa\mbox{}^{\xi}{\cal K}}{\pa z}$ into 
$\frac{\pa\mbox{}_{\eta}^{\xi}{\cal K}_{\Lambda\Sigma}}{\pa z}f^{\Lambda\Sigma}$ we obtain 
\be
\begin{array}{ll}
G_{\Gamma\Delta} & = {\cal G}_{\Gamma\Delta} + (1 - \mu){\cal R}_{\perp\perp}f_{\Gamma\Delta} 
+ f^{-\xi}
\left(
- \frac{\pa\mbox{}^{\xi}_{\eta}{{\cal K}^{\Lambda}}_{\Delta}}{\pa z}  
+ (\mu - \eta)  
\left[  
\frac{\pi}{1 - 4\eta}  \frac{\pa \mbox{}^{\xi}_{\eta}{\cal K}_{\Lambda\Sigma}}{\pa z}f^{\Lambda\Sigma} 
+ (1 - \pi)\frac{\pa\mbox{}^{\xi}{\cal K}}{\pa z} 
\right] 
f_{\Gamma\Delta}
\right) \\&
+ (1 -2\xi)                                                                                     {\cal K} {\cal K}_{\Gamma\Delta}
+ \left[2\xi\mu - \frac{1}{2} - \frac{2\pi\eta(\mu - \eta)}{1 - 4\eta}\right]                   {\cal K}^2f_{\Gamma\Delta} 
+  \left[\frac{1}{2} - \mu + \frac{2\pi\eta(\mu - \eta)}{1 - 4\eta}\right]                      {\cal K} \circ {\cal K}f_{\Gamma\Delta},
\end{array} 
\ee
which requires $\xi = \frac{1}{2}$, whereupon the two remaining equations become 
identical: $\mu - \frac{1}{2} = \frac{2\pi\eta(\mu - \eta)}{1 - 4\eta}$, which clearly 
has many solutions.  A particularly neat one is to take $\eta = 1$ so that the primary 
objects are `gravitational momenta' and $\pi = 1$ so that only two normal derivative terms 
appear in the `bulk' term.  Then $\mu = \frac{7}{10}$ so the BEFE's read 
\be
G_{\Gamma\Delta} = L_{\Gamma\Delta} + B_{\Gamma\Delta},
\ee
\be
L_{\Gamma\Delta} = {\cal T}_{\Gamma\Delta} + \frac{3{\cal T}_{\perp\perp} - {\cal T}}{10} f_{\Gamma\Delta} 
\mbox{ } , \mbox{ } \mbox{ }
B_{\Gamma\Delta} = \frac{1}{\sqrt f}\left( \frac{\pa {p^{\Pi}}_{\Gamma}  }{\pa z} f_{\Pi\Delta} - \frac{1}{10}\frac{\pa p_{\Lambda\Sigma}}{\pa z}f^{\Lambda\Sigma} f_{\Gamma\Delta} 
\right).
\ee

Of course, it would make sense to particularly investigate the difficult geometrical problem 
`$B_{\Gamma\Delta} = 0$ on some embedded hypersurface' for such $B_{\Gamma\Delta} = 0$ 
corresponding to no quadratic terms, since by the same arguments as above, such a model 
would be sure to give braneworld physics distinguishable from that hitherto studied.

\subsection{Two Further Comments about Building SMS-type Braneworlds}

First, so far we have talked in terms of the ${\cal Y}_{\Gamma\Delta} = 
T_{\Gamma\Delta} - \lambda f_{\Gamma\Delta}$ split of the matter contribution 
to relate our work as clearly as possible to its predecessors in the literature.
However, from the outset \cite{SMS} it was pointed out that this split is not unique.  
On these grounds we would prefer to work with the unique trace-tracefree 
split in which all the $\lambda f_{\Gamma\Delta}$ contributes to the trace part.  
The (4, 0) version of this split is used in Sec 6.  

Second, given a fixed type of bulk energy-momentum such as the ${\cal T}_{AB} = 0$ 
of NKK or the ${\cal T}_{AB} = \Lambda g_{AB}$, then 
establishing an embedding requires the existence of a suitable compensatory characterization 
of the bulk geometry.  The GR line of thought would be to only take results within 
such schemes seriously if they are robust to the addition of bulk matter fields. 
Of course privileged choices of bulk could arise from further theoretical
input.   We argue below that the theoretical arguments behind some privileged choices 
in the literature for the bulk energy-momentum are not convincing enough to anchor 
strongly credible physical predictions.  One would rather require rigorous and general 
theoretical input following directly from some fundamental theory such as string theory.  

The vacuum choice of bulk of the induced matter NKK approach is always possible 

\noindent given the 
premises of the CM result.  Then ${\cal R}_{AB} = 0$ is claimed to be a complete 
geometrization of matter \cite{Wesson}.  However, the CM result holds equally well for any 
other analytical functional form of the energy-momentum.  Furthermore, this approach considers 
only 1-component (`induced') matter; counting degrees of freedom shows that it cannot be 
extended to many important cases of fundamental matter. Whereas allowing for more extra 
dimensions could improve similar situations \cite{Wesson, Witten}, unification requires 
geometrization of the fundamental matter laws themselves, whilst this `induced matter' 
approach only geometrizes solutions of the EFE's coupled to matter of unspecified field 
dependence.\fn{For clarity, compare KK theory proper, in which the electromagnetic 
potential \it and Maxwell's equations \normalfont are geometrized, as the $f_{\Gamma z}$ 
portion of the 5-metric and the $G_{\Gamma z}$ equations of the KK split respectively, 
along with a scalar field.  NKK theory could geometrize a more general vector field than 
the electromagnetic potential in addition to the generalization of the KK scalar field that 
is all that is usually considered.  Now, this vector field is no known vector field 
of nature for its field equations does not in general correspond to that of any known vector 
field (although clearly the field equation contains the inhomogeneous vacuum Maxwell 
equations as a subcase since KK theory would be included in this way within NKK theory). 
For sure, this vector field (or even more obviously the scalar field) is not capable of 
being a simultaneous geometrization of the individual vector fields of the standard model.
If one were to treat these fields collectively, one throws away all that is gained by 
keeping their classical identities separate:  the results of Weinberg--Salam theory 
and QCD that are obtained then by quantization.}

The ${\cal T}_{AB} = \Lambda g_{AB}$ choice of bulk (an example of which is pure AdS) 
is clearly also always possible given the premises of a particular case of the generalized 
CM result.  
However, the motivation we wish to discuss is the argument for pure AdS bulks from string theory. 
This is not generally justifiable since firstly, 
bulk gravitons are permitted so the bulk geometry would generically contain gravity waves. 
Secondly, bulk scalars ought to be permitted since they occur along with the 
graviton in the closed string spectrum \cite{Polchinski}.  The content of the closed 
string spectrum thus places interesting restrictions on bulk matter rather than completely 
abolishing it.  From the perspective of 5--d GR, evidence for the stability of vacuum or AdS bulks 
(and of any resulting physical predictions) to the introduction of suitable bulk fields would 
constitute important necessary support for such models and their predictions.  

Finally note that use of arbitrary smooth bulk ${\cal T}_{AB}$ does not affect the form of the junction conditions since
only the thin matter sheet contribution to 
${\cal T}_{AB}$ enters these.  The conclusion of our second point is that there is no good reason not to explore at least certain kinds of bulk matter in order to have 
a more general feel for how these thin matter sheet models behave \cite{scalarbulk}.   

\subsection{Sideways York and Thin-Sandwich Schemes with Thin Matter Sheets}

First we explain how our criticism of (3, 1; 1) methods in general in Sec 2.4 are applicable 
to the case with thin matter sheets.  As regards the ``evolution" w.r.t $z$, the issue of 
causality holds regardless of whether thin matter sheets are present, and the issue of 
well-posedness not being known will become particularly relevant in the study of sufficiently 
general situations in which rough function spaces would 
become necessary to describe the evolution of thin matter sheets (see Sec 6.1).  
The Campbell--Magaard scheme is of limited use in models with thin matter sheets not only because analytic functions are 
undesirable (and definitely inapplicable to sufficiently general situations) 
but also because the junction condition imposes restrictions on ${\cal K}_{\Gamma\Delta}$ which prevents 
these being subdivided into the knowns and unknowns of Magaard's method.  These restrictions remain even 
if one considers suitably well-behaved thick matter sheet models.  
In the next paragraph we outline how much we expect can be achieved with the (3, 1) version of the 
York method applied on the thin matter sheet, but recall that even if this does provide data sets, one is next confronted with the difficulties of the (3, 1; 1) ``evolution" scheme.    
The main message is that the choice of methods which properly respect the difference between 
space and time is absolutely crucial.  Thus, although at the simplest level (3, 1; 1) methods which build 
higher-dimensional bulks about the privileged (3, 1) worlds may look tempting, general attempts at proceeding thus 
are hampered by the required mathematical tools simply not being available and by these attempts not being 
in accord with the usual notion of causality.
  
To date (4, 1) worlds built from (3, 1) ones have relied on very simple specialized ans\"{a}tze, such as 

\noindent A) $z$-symmetric surfaces ${\cal K}_{\Gamma\Delta} = 0$ with known metric $f_{\Gamma\Delta}$, 
whereupon the vacuum Codazzi equation is automatically satisfied and then 
$R = 0$ is required from the Gauss constraint.   

\noindent B) ${\cal K}_{\Gamma\Delta} = Cf_{\Gamma\Delta}$ with known $f_{\Gamma\Delta}$, 
for example to obtain the Randall--Sundrum bulk \cite{stdw2} or slightly more general solutions \cite{AL}.  
Now the maximal subcase ${\cal K} = 0$ of the CMC condition is a generalization of A), 
whereas the full CMC condition itself is a generalization of B).  Moreover, now the metric 
is to be treated as only known up to scale.  Thus one would generally only know the full 
metric of each model's (3, 1) world once the `wave Lichnerowicz equation' for the embedding 
of this world into the (4, 1) world is solved.  So one loses the hold from the outset 
on whether each model will turn out to contain an interesting (3, 1) world.  
Nevertheless, some of the (3, 1) worlds will turn out to be of interest.  Furthermore, one should 
question the sensibleness of any ideas involving the prescription of full 
(3, 1) metrics if the most general technique available fails to respect such a prescription.  This 
point is more significant for 
(3, 1) data than for (3, 0) data because conformally-related metrics 
have different non-null geodesics.  For (3, 0) data no physical significance 
is attached to spatial geodesics, but for (3, 1) there are timelike geodesics which are 
physically interpreted as paths of free motion of massive particles.  So a 
$(\tilde{M}, \tilde{h}_{ab})$ spacetime which is conformally related to $(M, h_{ab})$ 
is different physically (for example one could violate energy conditions the other one does 
not violate). One can get out of this problem by either attaching no physical significance 
to one's inspired guesses for $(M, h_{ab})$  or by hoping for unobservably tiny 
deviations between the geodesic curves of the two geometries.  

In the specific case of thin matter sheets, by the j.c's, A) implies that 
${\cal Y}_{\Gamma\Delta} = 0$, whilst  B) implies that ${\cal Y}$ is a hypersurface 
constant. The maximal condition implies that ${\cal Y} = 0$ 
whilst the CMC one implies that ${\cal Y}$ is a hypersurface constant.  
So whereas the maximal and CMC ans$\ddot{\mbox{a}}$tze are more general than 
A) and B) respectively, they are nevertheless restricted in this braneworld application.  
Notice, however, that a number of interesting cases are included: vacuum, radiative matter and electromagnetic matter 
are all among the ${\cal Y} = 0$ spacetimes.

It is important to note that unlike the usual GR application, the choice of a 
hypersurface to be a brane is not a choice of slicing because localized 
energy--momentum is to be pinned on 
it. Almost all reslicings would fail to isolate this energy-momentum 
on a single slice.  We know of no good reason why the brane should be CMC 
nor what value the CMC should take on it. However, at least this is a well-defined 
notion and it is substantially simpler to solve for than in general because of the decoupling 
of the Gauss and Codazzi constraints.

As regards possible use of either of the two forms of the thin sandwich conjecture
to treat branes, we first distinguish between thin sandwiches between 2 nearby branes
and thin sandwiches which have a brane on one side and an undistinguished hypersurface on the other.
One should be aware that non-intersection requirements may be different in these two cases,
and also different from that of the original TSC setting of 2 unprivileged spacelike
hypersurfaces.
Second, one would have to take ${\cal Y}_{\Gamma\Delta}$ as unknown until
it can be deduced from the ${\cal K}_{\Gamma\Delta}$ evaluated from the TS procedure.
Finally we caution that TSC schemes need not always exist.
They require the ``lapse'' to be algebraically-eliminable,
which for example is not the case for the analogous $\Phi$ of the KK split.

\section{Applying (r, 1; 1) Methods to Remove Singularities}

We have argued in Secs 2-3 in favour of (4, 0; --1) schemes over (3, 1; 1) schemes.  
However, it is often suggested that (3, 1; 1) schemes may be useful since 
in certain particular examples (3, 1) spacetime singularities are `removed' by embedding into 
nonsingular (4, 1) spacetimes \cite{sing1,sing2}.    
Whereas embeddings are undoubtedly useful tools in the study of (3, 1) 
spacetimes (for the purpose of algebraic classification \cite{mathemb, MC}), 
we ask what is the status of their use to remove cosmological singularities.  
Is it mathematically rigorous, generally applicable and physically meaningful?

By `singular' we mean geodesically-incomplete \cite{sclass}.    
In the study of singularities, timelike and null geodesics (t.g's and n.g's) are the only objects of fundamental importance since 
they are the curves privileged by the free travel of massive and massless particles respectively.
The other objects associated with the study of singularities arise as a matter of convenience rather than being fundamental.  

Such objects include the expansion and shear of geodesic congruences, and curvature scalars.  
Because these objects are closely-linked to material in this paper, we focus 
first on them (Sec 5.1), explaining how they arise in the study of singularities and then 
demonstrating how the embedding and embedded versions of these objects can be very different (Sec 5.2).  
We then discuss the more important issue of geodesic incompleteness in the 
context of embeddings (Sec 5.3).  An example is used to illustrate some of these points and others 
(Sec 5.4).  A lack of rigour in singularity-removal claims is uncovered in Sec 5.5.
We finally treat the case with thin matter sheets in Sec 5.6.  

\subsection{Secondary Objects in the study of Singularities}   

For ($q$, 1)-dimensional cosmology, smooth congruences of past-directed normal t.g's with normalized tangents denoted by $u^a$ are considered.    
The decomposition 
\be
\mbox{\sffamily B\normalfont}_{ab} \equiv D_au_b = \frac{\theta}{q}h_{ab} + \sigma_{ab} \mbox{ } \mbox{ } + \omega_{ab}     
\label{Bsplit}
\ee
provides the \it expansion \normalfont $\theta \equiv h^{ab}\mbox{\sffamily B\normalfont}_{ab}$ and \it shear \normalfont 
$\sigma_{ab} \equiv \mbox{\sffamily B\normalfont}_{(ab)}^{\mbox{\scriptsize T\normalsize}}$.  
The twist $\omega_{ab} =  \mbox{\sffamily B\normalfont}_{[ab]}$ is zero for the normal congruences considered here, 
in which case $\mbox{\sffamily B\normalfont}_{ab}$ is an extrinsic curvature $\theta_{ab}$.  
The corresponding normal Raychaudhuri equation [c.f (\ref{normRay})] can now be considered.    
Although for $\theta_0 > 0$ this would usually mean that a caustic develops, 
under certain global conditions a contradiction about the existence of conjugate points arises.  
Singularity theorems are thus obtained; in a cosmological context the 
simplest\fn{There are a number of other cosmological singularity theorems \cite{HE, Wald} not only 
because some have weaker assumptions, but also because one wants to be able to treat a number 
of pathological cases.  One such is the Milne universe:  
although this naively looks like a cosmology with its focusing of geodesics normal to 
$t =$const as $t \longrightarrow 0$, it is merely a region of Minkowski 
spacetime. So one would not want to include the Milne universe among the singular spacetimes. 
The way out of this is to demand that physically-meaningful cosmologies 
contain matter. This makes sense because the physical problem that may be 
associated with geodesics focusing is the possible pile-up of matter travelling along these 
geodesics leading to infinite densities. One succeeds in not including the Milne 
universe by use of singularity theorems hinging on the `timelike genericity condition' 
$R_{abcd}u^au^b \neq 0 \mbox{ }\mbox{ at least one point along each timelike geodesic}$, 
since the Milne universe is Riemann-flat.}  of these is (Hawking, theorem 1 of \cite{Hawking}). 

\noindent  For globally-hyperbolic (3, 1) GR spacetimes obeying the SEC and such that 
$\theta = C \geq  0$ on some smooth (spacelike) Cauchy surface $\Sigma$, 
then no past-directed timelike curve from $\Sigma$ can have greater 
length than $\frac{3}{|C|}$.  

\noindent By the definition of Cauchy surface, all past-directed t.g's are 
among these curves and are thus incomplete, so the spacetime is singular.  
Such theorems generalize to ($q$, 1) spacetimes (provided that these obey 
analogous energy conditions) in the obvious way since Raychaudhuri's 
equation clearly behaves in the same way for all $q$ and the global part 
of the arguments uses topological space methods that do not care about dimension.  

It must be noted that the singularity theorems are about existence whilst saying nothing about the nature of the singularity.    
Ellis and Schmidt began to classify singularities according to their properties \cite{sclass}, 
a difficult study which may never be completed \cite{Clarke}.   
Below we consider only the most elementary type of genuine singularities: \it curvature singularities\normalfont, 
for which at least one spacetime curvature scalar such as ${\cal R}$ or ${\cal R}_{AB}R^{AB}$ blows up.    

\subsection{Relating the Embedded and Embedding Secondary Objects}

Here we explain how knowledge of Gauss' hypersurface geometry renders it 
unsurprising that there are instances of singular spacetimes  being `embeddable' in nonsingular ones.  
This is because the behaviour of higher and lower dimensional objects used in the study of singularities clearly need not be related.    
As a first example, consider expansion.  In the normal case, $\theta_{\alpha\beta} = -\frac{1}{2\alpha}\frac{\pa e_{\alpha\beta}}{\pa t}$ and 
$\Theta_{ab} = -\frac{1}{2\alpha}\frac{\pa h_{ab}}{\pa t}$ from which follows 
\be
\theta = \frac{n - 1}{n}\Theta - \Theta^{\mbox{\scriptsize T\normalsize}}_{\perp\perp}.
\ee
So a blowup in  ${\theta}$ need not imply a blowup in $\Theta$.    
Thus the (4, 1) spacetime perspective on focusing of geodesics 
can be completely different from the perspective on some (3, 1) hypersurface.  
Figure 6 shows how one's (3, 1) notion of expansion generally corresponds 
to expansion and shear from the perspective of an embedding (4, 1) spacetime.       
In particular, what look like caustics or singularities in the (3, 1) spacetime could well correspond to 
no such feature in some surrounding (4, 1) spacetime. 

As a second example, consider curvature scalars, in particular the Ricci scalar.
Preliminarily, Gauss' outstanding theorem is a clear indication that extrinsic curvature can 
compensate for differences between higher and lower-dimensional intrinsic curvature.
 Our simple idea is to consider the implications for rigorous embedding mathematics of the 
generalization of this to the case where lower-dimensional curvature scalars become infinite. 
From the form (\ref{Agauss}) for the (3, 1; 1) Gauss constraint, 
\be
2\rho = 2{\cal R}_{\perp\perp} - {\cal R} = \frac{3}{4}{\cal K}^2 - {\cal K}^{\mbox{\scriptsize T\normalsize}} \circ {\cal K}^{\mbox{\scriptsize T\normalsize}}  - R
\label{singrem}
\ee 
clearly  - at least for some (3, 1) worlds which have Ricci-scalar curvature 
singularities $R \longrightarrow \infty$ - there will be surrounding (4, 1) worlds in which ${\cal R}$ 
(and the 5-d $\rho$) remain finite.  For it could be that 
${\cal K}^{\mbox{\scriptsize T\normalsize}} \circ {\cal K}^{\mbox{\scriptsize T\normalsize}} - \frac{3}{4}{\cal K}^2 \longrightarrow -\infty$ 
suffices to compensate for $R \longrightarrow \infty$. 
Thus it could be shear, expansion or both that dominate the compensation for $R \longrightarrow \infty$.  
If this involves $|{\cal K}| \longrightarrow \infty$, it means a (4, 0) caustic forms in the (4, 1) 
spacetime corresponding to the (3, 1) spacetime's singularity.  
Yet this is not the only nonsingular possibility; if $R \longrightarrow -\infty$  
it may be the shear that compensates, including cases in which  the blowup is pure shear.    

Finally, clearly the higher-dimensional singularity theorems hold so
even if one were to succeed in excising singularities from a lower-dimensional model, one would typically expect singularities to occur 
elsewhere in the resultant higher-dimensional models. The nature of these higher-dimensional singularities may not be the same as that of the 
excised lower-dimensional ones.  So, interestingly, by allowing extra dimensions, one would be even less certain of the character of 
singularities.

\subsection{Embeddings and Geodesic Incompleteness}

The fundamental importance of geodesics is problematic for embeddings since in general the (3, 1)-dimensional geodesics are not included among those  
of an embedding (4, 1)-dimensional spacetime.  There is then the dilemma of which of these sets of curves should be the physically privileged ones 
and thus be the set of curves whose extendibility is in question.  

If one wishes to postulate (4, 1)-dimensional GR, in addition to the (4, 1)-dimensional EFE's one must require that the matter follows the (4, 1)-dimensional geodesics.  
Then the issue of extendibility of the original (3, 1) geodesics becomes irrelevant since they are unprivileged curves in the (4, 1) spacetime.
But this is at the price of altering what the effective (3, 1)-dimensional physics is: often it will be of affine-metric type and thus in principle distinguishable from 
ordinary (3, 1) physics.  One cannot then necessarily claim that the recovered (3, 1) physics is as close to the usual (3, 1) physics as is widely claimed in the braneworld 
literature 

If these do not 
include the (3, 1) geodesics, then the (3, 1) question of what happens to 
these becomes physically irrelevant.  One would not then be extending one's (3, 1) geodesics but 
replacing them with distinct (4, 1)-dimensional geodesics, which might be viewed as unsatisfactory.  
There is the further complication in braneworld models that one might wish 
for (3, 1) and (4, 1) geodesics to both be relevant, as the paths of brane-bound and brane-unbound 
matter or gravity waves. It is then not clear what is meant by a singularity -- 
exactly which curves are supposed to be incomplete?  

We end with two loose possibilities about 
incompleteness which might sometimes be related: the (3, 1) singularities might 
be interpretable in terms of a (3, 1) hypersurface that becomes tangent to the characteristics 
of the bulk equations, and incompleteness might correspond to t.g's and n.g's 
being forced to exit the (3, 1) hypersurface. The second possibility is a variant on the theme 
of (4, 1) and (3, 1) geodesics whereby the (3, 1) geodesics are pieces of 
(4, 1) geodesics which are extendible only by replacing a piece of the original foliation with a new 
one that extends into what was originally regarded as the extra dimension.  

\subsection{Embedding Flat FLRW in Minkowski}

The following example illustrates many of the above points and leads us to further comments. 
Consider a (4, 1) spacetime with metric (see e.g \cite{sing1})
\be
g_{AB} = \mbox{diag}(g_{tt}, \mbox{ } e_{\alpha\beta}, \mbox{ } 
g_{zz}) = \mbox{diag}
\left(
-z^2, \mbox{ } t^{  \frac{2}{q}  }z^{  \frac{2}{1 - q}  }\delta_{\alpha\beta}, 
\mbox{ } \frac{q^2t^2}{(1 - q)^2}
\right) \mbox{ } , \mbox{ } \mbox{ } q > 1.
\label{funnymink}
\ee 
This is simple to treat because: 

\noindent 1) If we foliate it with constant $z$ hypersurfaces, a portion of each $z =$ const hypersurface has induced on it a flat FLRW cosmology.  
In particular, the coordinates have been chosen so that the $z = 1$ hypersurface is the FLRW cosmology with scale factor $t^{\frac{1}{q}}$. We restrict attention to 
$q \leq 3$ due to the DEC.  For $q \neq 2$ there is a (3, 1) Ricci scalar curvature singularity ($q = 2$ is the radiation universe, whence $R = S = 0$).  

\noindent 2) The (4, 1) spacetime is in fact Minkowski.

First, foliating the flat FLRW cosmology with constant $t$ surfaces,  
$\theta = \frac{3}{q t} \longrightarrow \infty$ as $t \longrightarrow 0$, so there is 

\noindent(3, 0; --1) focusing as the Big Bang is approached.  
Furthermore, \it only \normalfont focusing occurs: $\theta^{\mbox{\scriptsize T\normalsize}}_{\alpha\beta} = 0$.  Next foliate 
the (4, 1) spacetime with constant $t$ surfaces.  The (4, 1) spacetime contains many FLRW worlds on constant-$z$ surfaces.  Build a congruence 
by collecting the FLRW geodesics on each $z =$ const slice.  Then the (4, 0; --1) expansion ${\Theta} = \frac{q + 3}{ztq}$ also blows up as $t \longrightarrow \infty$, 
but there is also a blowup of the corresponding (4, 0; --1) shear:
\be
\Theta^{\mbox{\scriptsize T\normalsize}}_{ab} \equiv \frac{q - 1}{4qzt}\mbox{diag}(-e_{\alpha\beta}, \mbox{ } 3g_{zz}),
\ee
so we have case 2 of Fig 7.  Also, whilst still both $\theta \longrightarrow \infty$ and $\Theta \longrightarrow \infty$ in this case, we know that the former viewed 
from within the $z = 1$ (3, 1) hypersurface corresponds to a genuine (3, 1) singularity, whilst the latter in (4, 1) would be a mere caustic.  

Second, consider the $z = 1$ slice of the (4, 1) spacetime.  Here the (3, 1; 1) expansion and shear are 
\be
{\cal K} = \frac{q - 4}{q t} \mbox{ }, \mbox{ } 
{\cal K}^{\mbox{\scriptsize T\normalsize}}_{\Gamma\Delta} = -\frac{1}{4t}\mbox{diag}(3, \mbox{ } t^{\frac{2}{q}}\delta_{\gamma\delta}).
\ee
Thus for the physical range, both blow up as $t \longrightarrow 0$.  So the spacetime includes a $z = 0$, $t = 0$ point, at which there is a (4, 0) caustic.  
These blow-ups combine in ${\cal K}^{\mbox{\scriptsize T\normalsize}} \circ {\cal K}^{\mbox{\scriptsize T\normalsize}} - \frac{3}{4}{\cal K}^2$ to cancel $R$ for all values of 
$q$; for $q = 2$ the shear and expansion contributions exactly cancel each other.  The unphysical $q = 4$ case is an example of a pure shear blowup.

Although the congruence two paragraphs up is built to naturally include the geodesics of all the included FLRW universes, these turn out not to be the (4, 1) geodesics 
(nor pieces of them).  Rather, by considering the coordinate transformation between (\ref{funnymink}) and standard (4, 1)-dimensional 
Minkowsi coordinates, it is easy to show that the Minkowski geodesics pierce the $z =$ constant surfaces that are the FLRW universes 
(more significantly in the early universe than today).  As $t \longrightarrow 0$ the $z=$ constant surfaces approach the null cone so the example is an illustration of 
the foliating (3, 1) hypersurfaces becoming tangent to the characteristics at the point of interest.  Also the foliation breaks down as $t \longrightarrow 0$ 
because the family of hypersurfaces of constant $z$ intersect at $t = 0$.  All these points are 
illustrated in figure 6 b).                               

Finally, the example cannot be taken to be typical by the (4, 1) version of genericity.  
Indeed, we know that not many (3, 1) spacetimes can be even locally embedded into (4, 1) Minkowski \cite{MC}.  
The value of particular examples is limited to exhibiting \it possible \normalfont behaviours.  
One would require this to be upgraded to the study of  large classes of examples to assess \it probable \normalfont behaviours.  
Indeed, one would expect higher-dimensional cosmological models to exhibit a range of behaviours distinct from that of the Big Bang (including nonsingularness).  
This follows from knowledge of particular examples (from the algebraic study) of inhomogeneous cosmologies in standard GR \cite{Kras}.  
We especially note that if highly-symmetric cosmologies are haphazardly embedded, 
one would often expect the resulting higher-dimensional models to be less symmetric i.e 
more anisotropic and inhomogeneous than the original lower-dimensional models
[as naively suggested by the appearance of (4, 0; --1) shear from models containing no (3, 0; --1) shear].  
So the resulting higher-dimensional models are expected occasionally to exhibit more unusual behaviour than the lower-dimensional models from which they arose,  
and this includes the possibility of nonsingular cosmologies.  Indeed, we would expect that an increase of dimension increases the variety of possible 
behaviours in inhomogeneous cosmology.  But the bottom line in the study of 4-d and 5-d singularities is the same: one is ultimately interested 
in generic behaviour and this requires more sophisticated means of study than building individual particular examples.     

\subsection{Nonrigorousness of Singularity Removal by Embedding}

There are arguments against the use of embedding results toward making general statements about singularity removal.  
Our Secs 2.4 and 3 provide arguments against the use of the CM result (some of which also hold against the use of further embedding theorems).  
In particular, given a spacetime, there are so many possible embeddings.  Were nature higher-dimensional, why should it choose a particular 
nonsingular embedding out of an infinity of (nonsingular or singular) embeddings that mathematically exist?  The more extra dimensions are present, 
the more severe this nonuniqueness is.  This makes singularity removal by such embeddings physically-questionable.  
The CM scheme, and other embedding schemes such as the 10-d Minkowski embedding theorem \cite{mathemb}), are  
\it local\normalfont.  This localness includes the notion that the theorem is only applicable to a small region of the 
original manifold.  We question whether this need include the approach to a singularity, since these are edges of the lower-dimensional spacetime 
and thus have no neighbourhood.  \it Singularities are global features of spacetime\normalfont.  However even some of the global embedding theorems 
(which require very many extra dimensions) are stated for the analytic functions \cite{mathemb}.  But low differentiability may be typical 
in the approach to singularities \cite{Clarke}.  It becomes much harder to make any general statements once one accepts that spacetime is not analytic! 

One may view at least some (3, 1) singularities as projective effects due to taking a badly-behaved foliation (for example a foliation that becomes
tangent to the characteristics), manifested by the formation of caustics or pure shear blow-ups.  The question is why there is confusion in reversing this projection procedure 
to embed 

\noindent(3, 1) singularities into nonsingular (4, 1) spacetimes.  If an embedding of the ($r$, $s$; $\epsilon$) type such as those considered in this paper is to be used, 
it is mandatory that the region of hypersurface to be embedded be entirely of one signature $s$ (the CK theorem demands a 
\it nowhere-characteristic \normalfont lower-dimensional manifold).  Also, from the lower-dimensional perspective, the singularity cannot be included in the set of points on 
which data is prescribed because it is an edge of spacetime, not a point of spacetime.  Were one to try and include it by providing data `right up' to that edge, one would 
expect that the data would become badly-behaved e.g some components of ${\cal K}_{\Gamma\Delta}$ could be forced to be infinite.  
Under the various circumstances above, embedding theorems become inapplicable to reconstruct the higher-dimensional spacetime.

\subsection{(n, 1; 1) Singularity Removal and Thin Matter Sheets}

Finally, we have two comments about the application to thin matter sheets. 
First, there is less scope for singularity removal along the lines of (\ref{singrem}).  
For, as it is the energy-momentum on the brane, ${\cal Y}_{\Gamma\Delta}$ is presumably 
finite for a nonsingular (4, 1) world so ${\cal K}_{\Gamma\Delta}$ is finite and so 
cannot cause blowups in ${\cal K}^{\mbox{\scriptsize T\normalsize}}\circ{\cal K}^{\mbox{\scriptsize T\normalsize}} - \frac{3}{4}{\cal K}^2$.
However, blowups in this quantity 
can still occur if the inverse metric is badly-behaved (corresponding to $f$ = 0).
Note however that the application of the CK theorem to GR requires this not to be the case 
everywhere within the region of applicability.

Second, one would of course require the investigation of more elaborate 
(e.g `warped') embeddings than the example discussed above in order to 
investigate whether some embeddings lead to 5-d geodesics that exit the brane.
If the (3, 1)-dimensional geodesics on the  brane are not included 
among those of the (4, 1) dimensional bulk, it is not clear at all what is meant 
by `singular' since one then one would have to simultaneously consider some notion of extendibility
two congruences of privileged curves of different dimensionality.

Under the decomposition (\ref{ADMs}), the 5-d geodesic equation $\ddot{x}^A + {\Gamma^{A}}_{BC}\dot{x}^A\dot{x}^B = 0$ becomes 
\be
\ddot{x}^{\Gamma} + {\Gamma^{\Gamma}}_{\Delta\Sigma}\dot{x}^{\Delta}\dot{x}^{\Sigma} + 2{{\cal K}^{\Gamma}}_{\Delta} = 0 \mbox{ } ,  
\label{geo1}
\ee
\be
\ddot{x}^{\perp} + {\cal K}_{\Delta\Sigma}\dot{x}^{\Delta}\dot{x}^{\Sigma} = 0.
\label{geo2}
\ee
if one upholds the use of normal coordinates.  In order for (\ref{geo1}) to reduce in all cases to the (3, 1) geodesic equation on the brane, 
one requires everywhere on the brane $\dot{x}^{\perp} = 0$ or ${{\cal K}^{\Gamma}}_{\Delta} = 0$. By (\ref{geo2}), maintenance of $\dot{x}^{\perp} = 0$ 
along all geodesics is impossible unless ${\cal K}_{\Delta\Sigma} = 0$.  Thus in all cases ${\cal K}_{\Delta\Sigma} = 0$ is required on the brane.  
But this means that ${{\cal K}_{\Delta\Sigma}}^+ = {{\cal K}_{\Delta\Sigma}}^-$ or equivalently ${\cal Y}_{\Delta\Sigma} = 0$ i.e the absence of a brane.  Thus 
in SMS-type braneworlds the (3, 1) geodesics are not generally among the (4, 1) geodesics.  So if one wishes to adhere to the postulate that bulk-traversing fields 
follow (4, 1) geodesics and fields confined to the brane follow (3, 1) geodesics, one should accept that the restriction of the former to the brane generally does 
not coincide with the latter.  Thus there may be observable consequences in the (3, 1) physics:  freely-falling bodies composed of confined matter 
would follow distinct paths from freely-falling bodies composed of matter capable of traversing the bulk but which happens to be moving along within the brane, 
which represents a violation of the equivalence principle, and the causal structures within the brane due to confined photons and due to gravitons which happen to be 
moving along within the brane would generally be distinct.  
Also the above new conceptual difficulty about the definition of singularities is thus indeed relevant to brane cosmology.

\section{Application of (n, 0; --1) Methods to Thin Matter Sheets}
\subsection{Difficulties and Modelling Assumptions}

We are interested in the study of ($n$ + 1)-dimensional GR models with thin matter 
sheets using only well--studied, well--behaved mathematical 
techniques which also make good physical sense and are applicable to 
full EFE systems.
In our view a good way of achieving this is via the ($n$, 0) IVP, 
followed where possible by the heavily-protected ($n$, 0; --1) CP.  

We provide a hierarchy of modelling assumptions together with associated 
sources of difficulty, and argue that evolution is more problematic than data construction.  
We concentrate on the latter, making use of all the attractive features of the 
York method mentioned in Sec 3, albeit this elliptic mathematics is in a new 
setting (see Sec. 6.2-3). The differences come about due to the following difficulties of 
relevance to the hierarchy of modelling assumptions.     
Firstly, the roughness of function spaces required in the rigorous 
treatment of certain thin matter sheet models. 
Secondly, the requirement of novel asymptotics and boundary conditions (b.c's).     

Note that the increase in dimensionality itself does not affect much 
the data construction nor the study of well-posedness and stability of the GR CP, 
although there is some dimensional-dependence when the rougher function spaces come into use, 
due to the following two contrasting results.  

The first result is that in all dimensions, the presence of thin matter sheets means 
that the metric is rougher than $C^2$ (the functions with continuous second derivatives) 
because of the jump in the Riemann tensor at the sheet.  This corresponds to the metric 
being too rough to belong to the Sobolev class $H^4$.  

The second result is that the definition of a product\fn{This is a necessary complication 
because the EFE's are nonlinear.} for Sobolev spaces for the EFE's in harmonic coordinates according 
to Hughes, Kato and Marsden (HKM) \cite{HKM, Clarke} requires the Sobolev class of the metric 
to be no rougher than $H^{n + 1}$ [i.e $H^{4}$ for standard GR and $H^5$ for 
(4, 1)-dimensional GR]. Thus HKM's mathematics, the strongest used for both 
the GR IVP and CP in the review \cite{CBY}, is not generally powerful enough to deal with thin 
matter sheets.

More powerful mathematics is thus desirable, and more may well appear over 
the next few years (see the program starting with \cite{Klainerman}).  Realizations of part of the 
hierarchy of models below would be an interesting application for this sort 
of mathematics, which further motivates the study of increasingly rough Sobolev spaces, since the 
higher the dimension the greater the improvement to the established HKM mathematics 
is required for a full study of thin matter sheets.  We must also mention that in spaces 
rougher than $H^4$ there is no guarantee of geodesic uniqueness \cite{HE}, so complete 
conventional physical sensibleness is generally lost no matter what rougher Sobolev class 
results are proved. This shortcoming is clearly dimension-independent 
(from the dimension-independence of the form of the geodesic equation).  At least one need not 
go below $H^3$ whereupon geodesics need not even exist!  
 
The asymptotics problem \cite{AsADS} follows from supposing that 
`AdS' bulks  are desirable for string-inspired scenarios. The study of 
`AdS' bulks by relativists would entail the study 
of asymptotically AdS bulks, to permit a more general study of disturbances 
close to the thin matter sheet. The following comments are in order here.
First, in contrast to 
dS, AdS is infinite and hence possesses asymptotics. Second, whereas 
we will for the moment seek to avoid the function space difficulty, 
the asymptotics problem will remain 
in the `tractable scenarios'. In the usual GR, the use of AdS 
asymptotics for the application of the York method to small-scale astrophysics was neglected because of the 
negligible effect of a cosmological constant on such small scales and because of DOD arguments.
The braneworld application is however substantially different, so one might 
in this case have to study the York method with `AdS asymptotics'.  
This could affect the tractability of the Lichnerowicz equation, and also whether the existence of the crucial 
maximal or CMC slices is as commonplace for these new spacetimes as it was for the (3, 1) GR ones.
Furthermore, since one is interested in infinite planar branes, the 
asymptotics is \it directional\normalfont, `perpendicular' to the brane. 
However, in general there is no such notion of perpendicularity in GR.
Whereas one can locally define 
geodesics and draw hypersurfaces perpendicular to them, this procedure 
in general eventually breaks down, for example due to (spatial) caustic formation.  

We start our hierarchy by envisaging the most general situation possible for 
thin matter sheets within the framework of GR.  It is entirely legitimate to construct a (4, 0) initial 
data snapshot with whatever shape of thin matter sheet, but it is not legitimate 
to assume that any features of this are maintained over time unless it can be shown that the full 
evolution equations robustly maintain these features. Strictly, this is an 
\it initial-boundary problem\normalfont, such as occurs for water waves or for the 
surface of a star.  This is a very hard and quite new problem in GR \cite{RF, FN}.
Thus the evolution step is particularly hard both in its full generality and in the 
justifiability of simplifications such as a fixed boundary.  

Consider a thin matter sheet in an asymmetric (not $Z_2$ symmetric) bulk.  
One would ideally want to follow objects that crash into and possibly disrupt the thin matter sheet.  
One can then imagine asymmetric crashes which might punch though the thin 
matter sheet.  The sheet could thicken or disperse with time. It could emit a 
significant amount of gravitational radiation, which could moreover include gravitational shock 
waves that spread out the $H^3$ character that initially pertained only to the thin matter 
sheet and not to the smooth surrounding bulk.  That is, the $H^3$ character of a thin 
matter sheet data set $J$ could typically infect the whole of its causal future ${\cal J}^+(J)$ (Fig. 7). 
In simple words, why should the junction remain unscathed?  Why 
should the bulk in the immediate vicinity of the junction at later times be simple, 
smooth or known?  Addressing these questions goes beyond the reach of present-day techniques.  
Rather we note that some usually-tacit assumptions are used to definitely avoid these difficulties, 
and that we do not know whether any of these are justifiable in the 
physics governed by the GR evolution equations. One can presuppose a thin matter sheet 
exists at all times to prevent it being created or destroyed.  
One can allow only bulks with regular thin matter sheet boundaries to preclude shock waves. 
One could also preclude asymmetric crashes by presupposing $Z_2$ symmetry.

Even if discontinuous emissions are precluded, one can imagine smooth emissions 
and absorptions (symmetric or asymmetric) whereby the bulk interacts with the thin matter sheet.  
This would entail material leaking off or onto each side of the thin matter sheet.
This can be precluded by the presupposition that the thin matter sheet's energy-momentum 
resides at all times on the sheet, which is encapsulated by the well-known 
`equation of motion' of the thin matter sheet \cite{jns, MTW}. This is a significant
restriction on the dynamics of thin matter sheets in GR, often 
carried out in the name of tractability but which may not be realistic.  

In the Randall--Sundrum scenario, the brane can be placed anywhere in the bulk without affecting it.
We strongly suspect that this is not a typical feature in GR-based 
braneworlds, nor is it desirable since it dangerously marginalizes the ontological status of the bulk.
We take this feature to be too specialized to be included in the  
developments below.  
   
We close this section by briefly discussing modelling assumptions which avoid some of the above difficulties.  
We then implement these in two particular classes of tractable problems.  The desirability of 
each modelling assumption strategy and of other modelling assumptions from a 
string-theoretic perspective is discussed in Sec 7.      

I) The choice of $Z^2$ symmetry may well look and indeed be arbitrary from a 
GR point of view and indeed from a string-theoretic point of view (see Sec 7).
But without such a choice one simply cannot establish any specific  
embeddings if there are thin matter sheets (in the GR study of stars one requires the 
absence of surface layers to perform matchings). Thus one studies the restricted case with 
$Z^2$ symmetry, in which one integrates only up to the b.c provided by the j.c. 
In the IVP part one can assume whatever configuration of thin matter sheets. Thus this step is 
more justifiable than the subsequent evolution, which may require some of the above  
\it ad hoc \normalfont assumptions as to the existence and good behaviour of the thin matter 
sheet at all future times. So for the moment we are merely after the construction 
of (4, 0) initial data up to the junction. By themselves these data are useful 
in addressing issues such as the shape of the extensions of black hole horizons into the bulk, 
but the staticity and stability of these configurations remain unaddressed in the 
as-yet intractable evolutionary step of the problem \cite{401SS}.   

II) One could choose instead to work with thick matter sheets i.e ones that are finitely rather than infinitely thin \cite{Gregory}. Then the above troubles with the function 
spaces need not occur, since the walls would then not be rough in the above sense,  although they could still have the tractable level of roughness that is able to accommodate 
astrophysical objects just as in ordinary GR.  It is then easier to envisage the study of the evolutionary step of the thick matter sheet problem quite high up our hierarchy, 
since this more closely resembles the stellar surface problem.     
      
III) One could investigate closed shells rather than open sheets in order not to require directional asymptotics.  

\subsection{More Techniques for the (n, 0) IVP}

We cast the initial value constraints as  
\be
\triangle\phi - \mbox{Nonlin}(\phi) = 0 
\mbox{ } , \mbox{ } \mbox{ } 
\mbox{Nonlin} = - c\phi(R - M\phi^c + \mu^2\phi^b - 2\rho\phi^a) 
\mbox{ } , \mbox{ } \mbox{ }
M = \Theta^{\mbox{\scriptsize T\normalsize}} \circ \Theta^{\mbox{\scriptsize T\normalsize}}
\label{Nonlinlich}
\ee
\be
D_a\Theta^{\mbox{\scriptsize T\normalsize}ab} = j^b
\ee
\be
b = \frac{4}{n - 2} \mbox{ } , \mbox{ } \mbox{ } c = \frac{1}{(1 - n)b} \mbox{ } , \mbox{ } \mbox{ } a = \left\{ 
\begin{array}{ll} 
\mbox{arbitrary constant}     &     j^b = 0                 \\
\frac{1}{2c}                   &     j^b \neq 0.             \\  
\end{array} \right. 
\ee
We first consider the $j^b = 0$ case.  Then by definition $\Theta^{\mbox{\scriptsize T\normalsize}ab} = \Theta^{\mbox{\scriptsize TT\normalsize}ab}$ 
suffices in order to solve the Codazzi constraint.  Furthermore, in that case a particular scaling of $j^b$ is not required to keep the Codazzi constraint 
conformally-invariant, so preserving the DEC does not enforce a particular scaling of $\rho$.  So we are left to 
solve just the Lichnerowicz equation with the scaling of $\rho$ to be determined by our convenience.

We begin by considering the linearized Lichnerowicz equation. Set
$
\phi = \phi_0 + \varepsilon\phi_1$, 
$\rho = \rho_0 + \varepsilon\rho_1 
$
in (\ref{Nonlinlich}) and equate the $O(\varepsilon)$ terms to obtain   
\be
 [\triangle - \mbox{Lin}(\phi_0)] \phi_1 = \mbox{Inh}(\phi_0) 
\mbox{ } , \mbox{ } \mbox{ } 
\mbox{Lin}(\phi_0) = -c[R -(1 + c)M\phi^c_0 + \mu^2(1 + b)\phi^b_0 - 2(1 + a)\rho\phi_0^a ]
\mbox{ } , \mbox{ } \mbox{ }
\mbox{Inh}(\phi_0) = 2c\rho_1\phi^{1 + a}_0.
\label{linlich}
\ee
If asymptotically 
$
h_{ab} \longrightarrow h^{\mbox{\scriptsize A\normalsize}}_{ab}$ 
and $\phi_0 \longrightarrow \phi_0^{\mbox{\scriptsize A\normalsize}}, 
$
one obtains  
$[\triangle^{\mbox{\scriptsize A\normalsize}} - \mbox{Lin}(\phi_0^{\mbox{\scriptsize A\normalsize}})]\phi_1 = 
\mbox{Inh}(\phi_0^{\mbox{\scriptsize A\normalsize}})$ 
from which the asymptotic behaviour of $\phi$ for the full equation may be obtained.
Eq. (\ref{linlich}) is a linear elliptical equation\fn{Note however that 
$h^{ij}$ and $\Gamma^i$ have geometrical significance: our linear elliptic 
equation is on an underlying $n$-manifold $\Sigma$ (generally with boundary).  
We thus caution that if $\Sigma$ cannot be deformed to flat space, 
the interpolation methods between the $\Re^n$ Laplacian on a portion of $\Re^n$ and our operator 
cannot apply.  To avoid this trouble, one could study a sufficiently small 
region of $\Sigma$ (as $\Sigma$ is a manifold and hence locally $\Re^n$ or restrict the 
allowed topology of $\Sigma$ to be the same as that of a portion of $\Re^n$.}
\be
[h^{ij}\pa_i\pa_j  - \Gamma^i\pa_i - \mbox{Lin}(\phi_0)]\phi_1 = \mbox{Inh}(\phi_0)
\label{linell}
\ee
(where $\Gamma^i = h^{jk}{\Gamma^i}_{jk}$).  Many b.v.p's for such equations are well-studied \cite{CH, ellbooks, LU}.  
To give an idea of the techniques available, consider the Dirichlet problem: (\ref{linell}) in $\Sigma$ together with b.c  $\phi_1 = f$ on $\pa\Sigma$.     
On sufficiently-benign function spaces, the well-known uniqueness proof based on Green's identity holds, provided that Lin $\geq 0$ everywhere.  
This uniqueness can also be proved from the \it maximum principle\normalfont, the use of which applies to more general function spaces.  
One then proves existence from uniqueness \cite{LU}, by setting up a sequence of problems interpolating between the flat-space Laplace equation and (\ref{linell}).  
One can then interpolate between quasilinear elliptic equations such as the full Lichnerowicz equation and their linearizations.  
Then one has topological theorems about the fixed points of suitable maps, such as Leray--Schauder degree theory \cite{LU}, 
to prove existence of the solutions to the quasilinear elliptic equation's Dirichlet problem.  
Similar techniques are available for further b.v.p's \cite{LU}.  

Thus the chances of having a well behaved b.v.p for the Lichnerowicz equation are much enhanced
if the corresponding Lin $\geq 0$. To this end, notice that $M$ and 
$\mu^2$ are always positive whereas $R$ is of variable sign.  Contrary to what York assumes in the ordinary GR context, 
we must permit $\rho < 0$ since our application requires a negative bulk 
cosmological constant, which further complicates the analysis of which cases are guaranteed 
to be well-behaved.  There are three further simplifications one can try: 

\noindent i) Conformal flatness, which replaces $\triangle$ by the flat-space $\triangle_{\mbox{\scriptsize F\normalsize}}$ and wipes 
out the arbitrary-sign $R$ term.

\noindent ii) Maximality: $\mu = 0$ removes the highest-order term of Nonlin($\phi$).

\noindent iii) Setting $M = 0$. If this is combined with ii), one has time symmetry $\Theta_{ij} = 0$, which is regarded as overly 
simple in the standard applications of numerical relativity as it corresponds to the absence of gravitational momentum on 
the slice $\Sigma$.  

\noindent One can sometimes also use a trick: to scale $\rho$ like $M$, $R$ or $\mu^2$ and then fix it so as to 
cancel with this term.  Because we are interested in $\rho < 0$, only the first two of these possibilities are 
available.  

In the $j^a \neq 0$ case, the cancellation trick above is not available if one wishes the procedure to respect the DEC.  Simplification i) is convenient in this case.    
Whereas it is important to develop this case, it lies beyond the scope of this paper.

\subsection{The (4, 0) IVP with Thin Matter Sheets}

We work with the unphysical line element $ds^2 = dz^2 + e^{w}dx_{\alpha}dx^{\alpha}$ for some known trial function $w = w(r, z)$, 
so as to model a $S^2$-symmetric object such as a black hole or a star on the brane.  
For the $j^a= 0$ case, the IVP essentially reduces to the solution of the Lichnerowicz equation for some $\phi = \phi(r, z)$. 
We begin with the $T_{ab} = 0$ case of (the spatial projection of) (\ref{BWEM}).  
We first find the b.c's for the Lichnerowicz equation, then provide algorithms and then 
comment on the underlying analysis.  

Note that the b.c's are to be imposed on the physical metric 
and then written in terms of the unphysical quantities to have b.c's in terms of
the objects we start off with and the unknown conformal factor $\phi$.  
We begin with the inner conditions and then discuss asymptotic outer conditions.
To establish this sort of inner radial b.c, it is commonplace to use an isometry \cite{BY}.  For example, one could  
use $\tilde{h}_{ab}(x_{\alpha}) = \tilde{h}_{ab}(-x_{\alpha})$ 
$\Rightarrow$ 
$\left. \frac{\pa \tilde{h}_{ab}}{\pa x_{\alpha}}\right| _{x_{\alpha} = 0} = 0$ 
$\Rightarrow$ 
$\left. \frac{\pa\tilde{h}_{ab}(x_{\alpha})}{\pa r}\right|_{r = 0}$ leading component-by-component to 
the homogeneous Neumann b.c (the so-called reflection b.c) 
\be 
\left. \frac{\pa \phi}{\pa r}\right|_{r = 0} = 0,
\label{irbc}
\ee  
and the restriction $ \left. \frac{\pa w}{\pa r} \right| _{r = 0} = 0$ on the valid form of the known function.

One might worry in the case of black holes that $r = 0$ is singular.  For ordinary GR, one excises the singularity by use of  
an inversion-in-the-sphere isometry about some throat at radius $a$ of the black hole within the apparent horizon
(See Fig 8a).\fn{One has no choice but to work with apparent horizons rather than event horizons if one is given a single spatial slice.  At least the apparent horizon lies within the event 
horizon in GR.}  This gives an inner Robin condition 
\cite{BY} 
\be
\left.\left[\frac{\pa \phi }{\pa r} + \frac{1}{2a}\phi\right]\right|_{r = a} = 0.
\ee
For braneworld black holes one does not know how far the black hole extends into the bulk.  The main point of the study is to find this out (pancakes versus cigars).  
Our idea is then to guess a profile $r = f(z)$ along which to excise.  About each point on $r = f(z)$ an isometry in the corresponding 2-sphere may be applied (fig 8b), 
leading to an inner Robin condition
\be
\left.\left[\frac{\pa \phi }{\pa r} + \frac{1}{2f(z)}\phi\right]\right|_{r = f(z)} = 0.
\ee
The profile could be chosen so that it matches up smoothly with the $r = 0$ and $z = 1$ boundaries.  Once the problem is solved, one can find out whether this 
profile was a good choice or not (see fig 8c).  By picking a 1-parameter family of $r = f(z)$ curves, one could then iterate until a satisfactory profile is found.  

For the inner-$z$ ($z = 1$) b.c, we impose the j.c's (\ref{jcf}) and (\ref{prez2}), which now read 
\be
\tilde{e}_{\alpha\beta}^+ = \tilde{e}_{\alpha\beta}^-
\mbox{ } , \mbox{ } \mbox{ } \mbox{ }
\tilde{K}_{\alpha\beta}^+ - \tilde{K}_{\alpha\beta}^- = -\kappa_5^2\left(\tilde{Y}_{\alpha\beta} - \frac{\tilde{Y}}{2}\tilde{i}_{\alpha\beta}\right)
\label{scndjn}
\ee
(N.B these are imposed on the physical quantities).  Imposing $Z_2$ symmetry, the second of these becomes 
\be
\left. \left[\frac{\pa \phi}{\pa z} + \frac{\phi}{2}\left(   \frac{\pa w}{\pa z} - \frac{\kappa_5^2\tilde{\lambda}}{2}\phi \right)\right] \right| _{z = 1} = 0 
\label{nonlinbc}
\ee
by use of the definition of the physical $\tilde{K}_{\alpha\beta}$ in normal coordinates and then factorizing out the $e_{\alpha\beta}$.  This $Z_2$ 
isometry b.c is more 
complicated than usual because of the presence of the thin matter sheet.

That the $Z_2$ symmetry is termed a reflection appears to generate a 
certain amount of confusion.  This is because there is also the reflected wave notion of reflection, 
taken to give a Neumann b.c (such as in the $r = 0$ condition above). 
In the case of Shiromizu and Shibata (SS) \cite{401SS} (see also \cite{401N}), who furthermore treat 
our known $w$ as their unknown, and our unknown $\phi$ as the perfect AdS 
$\phi \propto \frac{1}{z}$, it has the nice feature that the other 2 terms in (\ref{nonlinbc}) cancel 
leaving a Neumann b.c 
\be
\left. \frac{\pa w}{\pa z} \right|_{z=1} = 0,
\label{funnyref}
\ee 
which might look like a reflection in both of the above senses.
However, this is only possible for $\phi \propto \frac{1}{z + q(r)}$ which may be somewhat restrictive.  Note 
also that (\ref{funnyref}) is an unusual notion of reflection in this context,
since it is of the conformally-untransformed (3, 0)-metric, whereas the natural notion of reflection isometry 
would be for the conformally-transformed (4, 0)-metric. 
Thus we conclude that the presence of the notion of reflection  (\ref{funnyref}) in SS's work is happenstance: it is 
separate from the $Z_2$ notion of reflection encapsulated in (\ref{scndjn}) and no 
analogue of it need exist in more general situations.  

Our b.c. (\ref{nonlinbc}) contains a reflection part $\frac{\pa\phi}{\pa z}$; 
the other part being a (nonlinear) absorption. This has the following implications.  First, the 
presence of absorption terms is interesting since pure reflection is the boundary condition of a perfect insulator in the potential theory of heat.  This suggests that pure 
reflection b.c's such as for pure AdS have built-in non-interaction of the bulk with its bounding brane, whereas our b.c may lead to (evolution models) with brane-bulk 
interactions.  Whereas neither the planar symmetry in (4, 1) spacetime of the underlying thin matter sheet nor the $S^2$ symmetry within the (3, 1) thin matter sheet of a 
simple astrophysical object generate gravity waves, we expect that spherical bumps on approximately planar branes in (4, 1) spacetimes to admit gravity waves, potentially 
giving rise to instabilities in the approximately AdS bulk braneworlds in which the branes contain astrophysical objects.  Our b.c could permit the modelling of such 
interactions.  Second the nonlinear absorption (much as arises in the theory of heat with temperature-dependent conductivity) is likely to complicate both analytic and 
numerical treatments of our b.v.p.

The first point above is one reason to favour our approach over that of SS (subject to overcoming the complications due to the second point).  
We now provide a further reason why the approach we present should be favoured on the long run.  
Whereas one could try to generalize SS's work by keeping the notion that $w$ be unknown and $\phi$ known but not $\frac{1}{z}$, whereupon our b.c (\ref{nonlinbc}) is 
interpreted as an inhomogeneous Neumann b.c in $w$, the main trouble with SS's work is that there is no good reason for their method to be directly generalizable 
away from its restrictive assumption $\Theta_{ij} = 0$.  
Whereas treating $w$ not $\phi$ as the unknown may 
give nicer b.c's, it is not tied to a method known to be amenable to less trivial solutions of the Codazzi equation than 
$\Theta^{\mbox{\scriptsize T\normalsize}}_{ij} = 0$.  Our proposed method seeks a workable extension to the significantly more general case  
$\Theta^{\mbox{\scriptsize T\normalsize}}_{ij} = \Theta^{\mbox{\scriptsize TT\normalsize}}_{ij}$, and speculatively to completely general 
$\Theta_{ij}$.    

Another nice feature in SS's work is that the matter is scaled so that the linearized p.d.e obtained is a Poisson equation.  
This p.d.e is simply invertible to obtain detailed asymptotics \cite{TanGarr, 401SS}.  For our scheme, the linearized equation (\ref{linlich}) is of Helmholtz-type rather than 
Poisson, complicating such a procedure.    We merely demand instead that $\phi \longrightarrow \frac{1}{z}$ for $z$ 
large wherever possible so that our models are `directionally asymptotically AdS'.  
As for large $r$, at least on the brane, one can impose asymptotic flatness as $r \longrightarrow \infty$.  It is less clear what one should prescribe in this respect off the 
brane.  A more detailed study of the asymptotics for our b.v.p's should be required as part of the actual construction of examples of data sets.\fn{Particularly because 
numerical integration is done on large but finite grids, subleading order asymptotics are helpful.}

Our proposed b.v.p is thus the mixed, nonlinear problem 
\be
\left\{
\begin{array}{l}
\triangle\phi = \mbox{Nonlin}(\phi) \mbox{ } , \mbox{ } \mbox{ } \\
\frac{\pa \phi}{\pa z} + \frac{\phi}{2}\left(\frac{\pa w}{\pa z} - \frac{\kappa_5^2\tilde{\lambda\phi}}{2}\right) = 0 \mbox{ for } z = 1 \mbox{ and } r \geq f(1) 
\mbox{ } ,  \mbox{ } \mbox{ }
\phi \longrightarrow \frac{1}{z} \mbox{ as } z \longrightarrow \infty 
\mbox{ } , \mbox{ } \mbox{ } 
\\\left. \frac{\pa \phi}{\pa r}\right. = 0 \mbox{ for } r = 0 \mbox{ and } z \geq f^{-1}(0) 
\mbox{ } , \mbox{ } \mbox{ } 
\phi \longrightarrow 1 \mbox{ } \mbox{as} \mbox{ } r \longrightarrow \infty
\\\frac{\pa \phi }{\pa r} + \frac{1}{2a}\phi = 0 \mbox{ on } r = f(z).
\end{array} \right.
\ee
One can use simplifications i) and/or ii) on this and still have a more general case than SS.  
Simplification i) might be plausible because 5-d AdS clearly admits conformally-flat spatial sections 
[$ds^2 = \frac{1}{z^2}(l^2dz^2 +dx_{\alpha}dx^{\alpha}$)] or maybe not, since it is regarded with suspicion in the (3, 1) GR 2-body 
problem \cite{BSrev}.  The trick at the end of Sec 6.2 can not be used to obtain particularly simple examples due to the 
following argument.  Unless one puts $\rho \propto \Lambda$ (constant) and has this maintained by non-scaling (i.e $a = 2$), it 
is overwhelmingly probable that the emergent $\tilde{\Lambda}$ is not constant.  Thus our freedom in $a$ is used up to ensure that $\tilde{\Lambda}$ is constant.  
Our algorithm for solving this problem is as follows:  

\noindent
i)   Prescribe the following unphysical quantities: the metric $h_{ab}$ and matter density $\rho$.    

\noindent
ii)  Pick any suitable $\Theta^{\mbox{\scriptsize TT\normalsize}ij}$ to solve the Codazzi constraint.  It follows that $M$ is known.     

\noindent
iii) Thus we can attempt to solve our b.v.p for the Lichnerowicz equation to obtain $\phi$.    

\noindent
iv) Then we can compute the physical bulk metric $\tilde{h}_{ab}$ and induced brane metric $\tilde{e}_{\alpha\beta}$ of our snapshot.

This assumes that we have correctly guessed the profile.  One would now check whether this is the case by solving for the apparent horizon.  
If this is unsatisfactory (fig 8c) then one would repeat the algorithm with adjusted profile.  

For nonvacuum branes ($T_{ab} \neq 0$), a similar argument enforcing $a = 2$ holds. The j.c may now be split into a trace b.c 
part,

\noindent
\be
\left.
\left[
\frac{\pa \phi}{\pa z} + \frac{\phi}{2e}
\left(
\frac{ \pa e}{\pa z} + \frac{\kappa_5^2Y\phi}{6}
\right)
\right]
\right|
_{z = 1} = 0
\ee
and a restriction on the tracefree part of the matter on the brane, which is 
\be
Y_{\alpha\beta}^{\mbox{\scriptsize T\normalsize}} = 0
\label{95}
\ee
for the metric ansatz and coordinate choice used.  
Our algorithm now becomes

\noindent
i)  One now requires the prescription of $h_{ac}$ everywhere and of $Y$.    

\noindent
ii) Pick any suitable $\Theta^{\mbox{\scriptsize TT\normalsize}ij}$ and hence $M$.     

\noindent
iii)Attempt to solve our b.v.p for the Lichnerowicz equation.   

\noindent
iv) Then we can compute $\tilde{h}_{ab}$ and $\tilde{i}_{\alpha\beta}$.  In this particular case, the simultaneous imposition 
of normal coordinates and an isotropic line element forces the restriction (\ref{95}) on the braneworld matter.  This is because one is applying too many coordinate restrictions; 
in this light the habitual practice of working in braneworlds using normal 
coordinates may be seen as a poor choice of gauge. If the above coordinate choices are not simultaneously 
made, the algorithm would contain further nontrivial equations [in place of 
(\ref{95})] to solve before the braneworld matter content can be deduced.

We assume that this nonvacuum application is for a star or clump 
of dust on the brane in which case there is no need to excise a corner with a profile as 
was done above to deal with the black hole singularity.  

Notice the lack of control of the physical metrics characteristic of theoretical numerical relativity.  
Our non-scaling of the matter at least gives visible control over the matter.  
The sensibleness of doing this is in fact tied to the negativity of the bulk $\rho$:
for $\rho \geq 0$ and $a \geq -1$, Lin receives a negative contribution with its 
tendency to encourage ill-posedness, whereas for predominantly 
$\rho \leq 0$, the danger \cite{CBY} lies in $a \leq -1$. 

A simpler example of b.v.p set up along the lines we suggest is that of Nakamura, Nakao and Mishima \cite{401N}.  
Their metric is cylindrical not spherical and they have a slightly simpler version of (\ref{nonlinbc}).  
Their case is still time-symmetric.  The linearized equation is then precisely Helmholtz, facilitating its inversion and the consequent more detailed 
knowledge of the asymptotics.

We also propose the 2-brane version of the above, in which the 
$z \longrightarrow \infty$ condition is replaced by another `parallel' 
brane boundary at $z = 1 + l$. This problem is close to that 
considered by Piran and Sorkin \cite{401lit2}.  In contrast however, their study involves
a periodically-identified fifth dimension to which there does correspond an isometry-based reflection Neumann condition $\left. \frac{\pa\phi} {\pa z} \right|_{z = 1} = 0$.  

We can provide a local (in function space) uniqueness proof for our proposed type of b.v.p,  
at the same level as that for standard GR black hole data in \cite{BY}.  In the conformally-flat case, 
suppose that $\psi_1 = \psi_2 + u$ for $u$ small.  Then the homogeneous linearized b.v.p in $u$ is applicable.  
We then have 
\be
\int_{\Sigma}(|\nabla u|^2 + \mbox{Lin}u^2)d^4x = \int_{\Sigma} (|\nabla u|^2 + u\triangle u)d^4x =  \oint_{\pa\Sigma} u\frac{\pa u}{\pa n}dS =  
\begin{array}{l}
\int_{z = Z} u\frac{\pa u}{\pa z}dS + \int_{r = R} u\frac{\pa u}{\pa r}dS \\
 - \int_{z = 1 + \epsilon} u\frac{\pa u}{\pa z}dS 
- \int_{r = \eta} u\frac{\pa u}{\pa r}dS 
\end{array}
\ee
by Green's theorem.  For $u \longrightarrow 0$ at least as fast as $\frac{1}{z}$, the first surface integral tends to zero.  
For $u \longrightarrow 0$ at least as fast as $\frac{1}{r}$  the second surface integral tends to zero.  
The fourth integral is zero by the Neumann reflection b.c.  Upon applying the linearization of the nonlinear b.c (\ref{nonlinbc}),
\be
\left.
\left[
\frac{\pa u}{\pa z} + \frac{u}{2}
\left(
\frac{\pa w}{\pa z} - \kappa_5\tilde\lambda\phi_2
\right)
\right]
\right|
_{z=1} = 0, 
\label{linrob}
\ee
the third surface integral is non-positive provided that the restriction $\frac{\pa w}{\pa z} \geq 0$ holds.  
Thus since Lin $\geq 0$ u must be zero.  For the non-conformally flat case, this argument requires $R \geq 0$.
The argument is not seriously changed in the black hole case when a corner is excised by a profile with a Robin condition on it.

Stronger existence and uniqueness proofs for our proposed problems are complicated by four factors:    

\noindent unboundedness, (albeit of the most benign kind), 
the boundary having corners,
mixed b.c's (i.e a piecewise prescription on the boundary)
and a portion of these being nonlinear.
The first is however of the simplest kind and the second and third are commonplace.  
The last is less usual but perhaps not so bad either because by 
$\triangle \phi = \frac{1}{\sqrt{\mbox{\scriptsize \normalsize}h}}\frac{\pa} {\pa x_i}\left(\sqrt{\mbox{}h}h^{ij}\frac{\pa\phi}{\pa x_j}\right)$
the problem can be written in divergence form, for which b.v.p's with nonlinear b.c's  are treated in Ch. 10.2 of \cite{LU}.    
The trouble is that the treatment there, unlike here, is neither for mixed b.c's nor for an unbounded region with corners.  
Thus, the strongest point we can make at present
is that there is good hope of obtaining existence and uniqueness 
results even for quite rough function spaces by more-or-less 
conventional, entirely rigorous mathematics for our proposed problem.  
This should be contrasted with the hopeless state of affairs with 
sideways prescriptions!  Once one begins to get good numerical results, it 
becomes worthwhile to explicitly prove the well-posedness of the method used, in order to support those results and the 
ongoing production of more such results. In this particular case, 
these results would serve to support and understand the crucial numerical step iii) in the above algorithms. 

To permit the even greater generality required to have momentum 
flows ($j^a \neq 0$), by our route one is forced to forfeit the DEC control over the matter.  
This is because at least the $\Lambda$ part of $\rho$ cannot scale in accord with the scaling of $j^a$.
We leave this further development, which additionally requires posing and solving 
the b.v.p following from the Codazzi equation, for a future study.

\subsection{The (4, 0) IVP with Thick Matter Sheets}

Many of the difficulties with evolution discussed above stem from the thinness of the matter sheets causing function-space-related problems.  But ordinary physics 
should not be sensitive to which function space is used.  One would hope that models with thick matter sheets would be more amenable to study, 
be good approximations to the thin matter sheets and in any case could be closer to reality than thin matter sheets.  

Bonjour, Charmousis and Gregory have used the (2, 1; 1) version of the general split
(\ref{gauss}), (\ref{cod}), (\ref{evK}) with scalar matter and consider 
both thick walls and the thin-wall limit
perturbatively in the presence of gravity. 
They further specialize to the case of a spherical domain wall, which they show collapses.  
It would be worthwhile if this sort of example, which combines the modelling assumptions II) and III), 
is investigated using the 

\noindent($n$, 0; --1) split subject to flat and to AdS asymptotics.  

For the particular example mentioned above, the matter profile is sigmoidal.
However, to approximate a thin brane, one would want instead a hump profile, 
as indicated in Fig 9a.  One could still use an inner b.c at $z = 1$, where now $K_{ij} = 0$ so that the linear Robin b.c
\be
\left. 
\left[
\frac{\pa \phi}{\pa \lambda} + \frac{\pa w}{\pa z}\frac{\phi}{2}
\right]
\right|
_{z = 1} = 0
\label{thickrob}
\ee
holds. 
One could in fact consider 2 problems (see Fig 9b): the flat sheet with directional asymptotics and the spherical sheet with a suitable inversion-in-the-sphere in place of 
the above inner b.c.  
Both of these problems avoid the main obstacles to existence and uniqueness proofs by 
possessing linear b.c's. The second problem is also invertible to an inner problem 
which is on a bounded region with no corners, for which results in \cite{LU} apply.  

These problems are for `plain' thick branes as opposed to branes containing spherically-symmetric objects. 
The first problem ought to more easily admit this extension.  
One might wish to study `plain' thin branes too, e.g in the context of branes in relative motion.  

Finally, suppose instead one attempted to use Magaard's method to construct (4, 0) thin-matter-sheet data.  
Then one benefits from $s = 0$ and the identification of $x_1 = const$ with the $z = 1$ brane prescribes the topology, 
removes some of the sources of nonuniqueness and the brane's energy-momentum endows physical significance upon the (3, 0) 
covariance of the elimination procedure.  In the thin matter sheet case, one is blocked by the nonanalyticity of $\rho$, 
but at least some thick matter sheet models could be built in this way. 

\section{Discussion: Modelling Assumptions and Stringy Features}

We first address \bf Q2 \normalfont by discussing how the work in this paper may be generalized and which of these generalizations may be regarded as stringy.  
We end with the conclusions of this paper as regards \bf Q1 \normalfont. 

We did not consider the possibility of a higher codimension \cite{cod2} which could substantially alter the nature of the embeddings used.  
For models with 2 or more times, it is simply not possible to stick entirely to our suggestion to build on established mathematical physics.  
Finally, our study could be complicated by having more than the Einstein tensor in the bulk.  
For example, one can have a nontrivial Lovelock tensor term \cite{Lovelock} in dimension 5, which arises from the variation 
of a Gauss--Bonnet term in the action, or one could have genuine higher-derivative terms \cite{hdt}.  
Whilst our incipient results such as the correspondence between ($n$, 0; --1) and 

\noindent($n$ - 1, 1; 1) schemes at the simplest level or our BEFE ambiguity 
will have counterparts in these theories,  these theories' equations are much larger than the EFE's, and their CP/IVP is much less well-studied than that of GR.  
Higher derivatives or non-standard bulk matter might lead to the violation of the energy conditions used in this paper.
  
We next discuss which features of GR-based braneworld models are desirable for string theory.  
First for some caveats.  We introduced the SMS braneworld as a GR generalization of the Randall--Sundrum scenario, moreover one that was as yet not sufficiently general 
for the purposes of GR in which the nature and existence of thin matter sheets at all times might not be a representative presupposition.  It could be that it is other aspects 
of the Randall--Sundrum scenario that are of interest in string theory since it also has particle physics aspects to it and because it is a toy model of Ho\v{r}ava--Witten theory.  
One may prefer to consider only such stringy scenarios rather than Randall--Sundrum or SMS.    
Also a great source of tension between string theorists and 
relativists is that the latter believe GR is suggestive of the importance of background--independent theories, 
whilst there is as yet no background--independent formulation of string theory available.    
Such ideas as conserved charges, flat sheets and simple bulks arise from assuming Minkowski or AdS backgrounds.  
Since GR is background-independent, one is interested there in generic solutions to the full field equations.  
This makes the notion of asymptotics necessary and subtle, and not necessarily compatible with flat sheets nor conserved charges.  
From a general perspective it is doubtful whether highly-specialized solutions such as 
Minkowski or AdS are likely to reflect the gravitational physics of the universe, unless it can be shown that large classes 
of generic solutions behave likewise or are attracted toward such solutions.  

String theory might favour many extra dimensions, although it is not clear how many of these should be large and how many compactified.  
Among the corrections to the Einstein tensor mentioned above, string theorists might favour the presence of the Gauss--Bonnet 
correction, since this is the only first-order correction for a heterotic string \cite{GS}; an infinite series of correction terms 
is predicted.   
String theorists might favour very flat walls on the grounds that these could be stabilized by being the lowest-energy carriers of BPS charge \cite{Polchinski, Turok}.  
 These arguments would rule against the use of highly symmetric bulk manifolds/orbifolds, although string theorists might try to invoke them as fixed backgrounds so as to 
perturb about them.    There may be different reasons for stability in GR and string theory.  For example, in contrast to string-theoretic arguments favouring 
reflection-symmetric orbifolds, a GR domain wall could be stable precisely because the (3, 1) bulk is different on each side, as the wall separates domains 
in which symmetry is differently broken. 
On the whole, the string theorist would agree that bulks could be complicated by the presence of bulk matter from the closed string multiplet, as   
mentioned in Sec 4.5.  At least in some models, gauge fields would be expected to occur only as fields confined to branes since they belong to the open string multiplet 
and some branes are where open strings end \cite{Polchinski}.   
As strictly, the branes are of Planckian thickness, so it is not a disaster that thick branes are 
favourable toward rigorous mathematics for GR-based scenarios (this is recycling the argument in footnote 11).    
Of course, one could worry that they are thin enough for quantum gravitational effects to be important. 

We now address \bf Q1\normalfont.  We feel that it is secure to say that GR-based braneworld 
models only make sense if realistic examples of them are stable to the dynamics which governs 
GR. We have investigated two schemes within which this study might be furthered: 
the (3, 1; 1) and (4, 0; --1) schemes.

As regards (3, 1; 1) schemes, before pressing on with something like the SMS braneworld by 
perturbation theory and dynamical systems analysis, we think that it is important to check 
that the SMS braneworld's interpretation is properly understood.  We have demonstrated that 
there exist many reformulations of SMS's BEFE's, thus making it clear that it is a choice as 
to whether the BEFE's contain SMS's Weyl term (requiring closure as some complicated 
third-order system such as in SMS's paper), and that it is a separate choice as to whether any 
terms homogeneously quadratic in the thin matter sheet energy-momentum tensor feature in the 
BEFE's. We have used this to illustrate the danger in truncating the BEFE's by setting bulk 
terms to be zero or to take a particular form.  The full (4, 1)-dimensional picture of the brane 
is required.  Building this picture by sideways evolution appears unlikely to be well-behaved 
mathematically or to make sense from a physical (causal) point of view.  We hope that this paper will 
either stimulate new research where these concepts are carefully thought about or force an 
admission that it is preferable to use a (4, 0; --1) scheme.  We have argued that the 
removal of singularities by the (3, 1; 1) scheme is unlikely to be generally possible
in a meaningful way.  Whichever of the two schemes one works in, one is well-advised to heed 
the old, careful literature about solving the Gauss--Codazzi system rather than starting afresh.  
We have carefully detailed which parts of this traditionally ($n$, 0) scheme easily pass over 
to the ($r$, 1) case.    

As regards the (4, 0; --1) scheme that we have argued is favourable, we have argued that its 
IVP step is cleaner than its CP step and by itself serves to answer a number of questions.  
We have provided a survey of obstacles and ideas to extend this IVP with a thin matter sheet 
away from the relatively trivial time-symmetric cases hitherto considered.  We also suggest 
that thick branes are more easily treated rigorously and that this choice should not affect 
the underlying physics if it is respectably robust.  We feel that the next stage should be to 
solve for such data numerically.  As this field grows, the underlying analytic mathematics will 
deserve detailed attention.  

We note that the scope of the (4, 0; --1) scheme would go as far as the study of colliding 
black holes, which are expected to be an important source of gravity waves.  One requires the 
kind of methods that we promote in this paper for GR-based braneworlds for the full study of 
such high-curvature processes.    Whilst this study is a tall order even for the usual GR 
\cite{BSrev, Cook}, these calculations may be required if braneworld effects can be large 
enough to substantially alter the extreme physics of colliding black holes.    

Finally, we summarize the aspects of \bf Q2 \normalfont that should be asked as the partial 
answers to \bf Q1 \normalfont begin to be built up.  Does the introduction of stringy features 
change the outcome to any of issues?  And so, are GR-based braneworlds adequate or typical 
frameworks for string theory? What scenarios should one explore in order to furnish string 
theory with reasonably unique predictions?
\\

\noindent \Large\bf Acknowledgements\normalfont\normalsize
\\

EA is supported by PPARC.  
We acknowledge useful discussions on a wide range of topics with 
Julian Barbour, Henk van Elst, James Lidsey, Roy Maartens, Malcolm MacCallum, Marc Mars, Nikolaos Mavromatos, 
Niall \'{O} Murchadha and Carlos Romero.

\mbox{ }

\mbox{ }

\mbox{ }

\mbox{ }

\mbox{ }

\mbox{ }

\mbox{ }

\mbox{ }

\mbox{ }

\mbox{ }

\mbox{ }

\mbox{ }

\mbox{ }

\mbox{ }

\mbox{ }

\Large \bf Figure captions\normalfont\normalsize
\vskip 0.3in
\scriptsize

\bf Figure 1\normalfont

\noindent a) For unspecified or general signature, we use the following notation. Let $(M, g_{AB})$ be a ($n$ + 1)-dimensional spacetime.
Let $\upsilon_0$ be an embedded spacelike hypersurface at parameter value $\lambda_0$, with unit normal $n^A$.  
The induced metric on $\Sigma$ is $h_{ab} = g_{ab} - \epsilon n_{a}n_{b}$.  
Consider another nearby spacelike hypersurface 
$\Upsilon_1$ at parameter value $\lambda_1$.  Then nearby slices fit together in accordance with the figure, 
where the lapse is $\alpha = \frac{1}{\sqrt{|g^{00}|}}$ and the shift is $\beta_i = g_{0i}$.  
In the particular case of a ($n$, 0;--1) split, we denote $\Upsilon$ by $\Sigma$, $n^A$ by $t^A$ and $\lambda$ by $t$
whilst for a ($n$ - 1, 1; 1) split we denote we denote $\Upsilon$ by $\Pi$, $n^A$ by $z^A$ and $\lambda$ by $z$.  

\noindent b) When we simultaneously talk of several splits in this paper, we use the notation presented in this figure.  
The white arrows represent the extrinsic curvatures associated with each of the embeddings.

\bf Figure 2\normalfont
 
\noindent A general means of study of hyperbolic p.d.e's is to require that the data be prescribed on spacelike surfaces [($n$, 0; --1) scheme] a), not on timelike surfaces 
[($r$, 1, 1) scheme]. 
b).  The line $J$ in a) represents the inclusion of a snapshot of a thin matter sheet, whilst the whole of the surface in b) 
could be the worldsheet of an arbitrarily thin matter sheet.    
 
\bf Figure 3 \normalfont (Domain of dependence property)  
\noindent We should not claim to know too great a portion of 
the future.  a) Given data for a hyperbolic system on a piece of a spacelike surface, 
we can predict the future only in the forward wedge (domain of dependence) within which all causal curves, 
i.e allowed paths of information flow, emanate from the data.  Outside this wedge, for all we know, large disturbances 
(represented in the picture by gravity waves) could influence the future arbitrarily soon.  

\noindent b) If we claim to know the 4-dimensional DOD, we preclude the influence by the extra dimension, thus making it 
largely redundant.  The 5-dimensional DOD of a thin sheet is small because of the thinness of the sheet.  Thus gravity waves 
could be lurking very close in the bulk.  This corresponds to the 4-d branewold EFE system not being closed.

\bf Figure 4\normalfont

\noindent a) The bucket-shaped construction that motivates the use of Sobolev spaces for the Cauchy problem for the 
Klein--Gordon equation.  

\noindent b) The analogous construction for the sideways Cauchy problem, which we argue is faulty on several counts.  
\it Sobolev spaces do not appear to be appropriate for the sideways Cauchy problem \normalfont.  
  
\bf Figure 5\normalfont

\noindent a) $\Pi$ is a ($r$, 1) (i.e timelike) hypersurface containing a data strip $T$ 
built by the Magaard method from prescribed `data for the data' on a timelike 
$x_1 = 0$. The shaded region $\Omega$ is the region of the ($r + 1$, 1) spacetime supposedly entirely controlled by $T$ 
i.e for which the Campbell method can validly provide the evolution.  Typically $T$ will not cover all of $\Pi$ 
in the $x_1$-direction, so there will be nearby points like $p$ outside $T$.  Consider the future light-cone 
with apex $p$.  No matter how thin in the $z$-direction one considers $\Omega$ to be, it is always pierced by causal curves 
such as $\gamma$ on or in the light-cone, by taking $\gamma$  to be at a sufficiently slender angle $\phi$ to $\Pi$.  
Thus information can leak into $\Omega$ from elsewhere than $P$, which is a contradiction.  Therefore, in parts at least, 
$\Omega$ is arbitrarily thin in the $z$-direction.  
One can envisage that sometimes there will be enough points like $p$ that there is no region $\Omega$ at all.  

\noindent b) Suppose $x_1 = 0$ is spacelike and is a section of the entire hypersurface 
$\Pi$ and $T^{\prime}$ is the data strip built from it by the Magaard method.  
There is now no room for points like $p$! However, typically $T^{\prime}$ will not cover all of $\Pi$ in the 
$x_1$-direction, so there will be nearby points like $q$ in the causal past $J^-(P^{\prime})$ not in $P^{\prime}$.  
No matter how thin one considers $\Omega$ to be, it is pierced by causal curves $\gamma^{\prime}$ from $q$ \it 
which do not pass through $P^{\prime}$ \normalfont, by taking these to be at a sufficiently slender angle $\phi^{\prime}$.  
So again, in parts at least, $\Omega$ is arbitrarily thin and sometimes there will be enough points like $q$ 
that there is no region $\Omega$ at all. Small pieces of timelike hypersurfaces need not hold useful information for 
hyperbolic-type systems!

\bf Figure 6\normalfont

\noindent a) A simple illustration of how 4-dimensional expansion may be interpreted as being mainly due to shear in a 5-dimensional embedding manifold.  

b) Diagram of the embedding of flat FLRW universes into Minkowski spacetime in the standard 5-d Minkowski coordinates $(T, X_1, X_2, X_3, Z)$.  
The curved surfaces are the FLRW spacetimes.  As one approaches $T = 0$ (corresponding to $t = 0$ in FLRW coordinates), each of these surfaces becomes 
tangent to the light cone (characteristics of Minkowski).  The foliation by these surfaces also becomes bad here because the surfaces intersect.  
Note also that the FLRW geodesics move within each of the surfaces whereas the Minkowski geodesics clearly pierce these surfaces.  Thus the 4-d and 
5-d geodesics in this example are not the same.

\bf Figure 7\normalfont

\noindent The spread of the $H^3$ region of a snapshot $\Sigma_L \bigcap J \bigcap \Sigma_R$ containing the $H^3$ junction $J$ 
between the 2 spacelike bulk pieces $\Sigma_L$ and $\Sigma_R$ upon which the prescribed data is smooth or analytic.  Then the 
evolutions in the domains of dependence ${\cal D}^+(\Sigma_L)$ and ${\cal D}^+(\Sigma_R)$ are smooth or analytic, but the causal future of 
$J$ will typically be $H^3$.  

\bf{Figure 8}\normalfont

\noindent a) Excision of the inner region in (3, 0) black hole data construction by an inversion-in-the-sphere isometry identification between two copies of the black hole.  
This is along the lines of the method of images in electrostatics.    

\noindent b) Excision of the inner region in (4, 0) braneworld black hole data construction would involve making a guess $r = f(z)$ for the excision region and then performing 
inversion-in-the-sphere isometries pointwise to identify two copies of the black hole.      

\noindent c) One is then to numerically solve for the data.  One can then find the apparent horizon.  If this intersects $r = f(z)$ (horizon 1) then the data 
may be bad in that the part of the shaded portion of the excised region may be in causal contact with its surroundings.  If the apparent horizon extends too far past 
the end of the excision region (horizon 2), one might worry that the singularity also extends past the excised region in which case some of the data prescribed along 
$r = 0$ should not have been prescribed since it lies on the singularity.  

\bf Figure 9 \normalfont

\noindent a) The metric, extrinsic curvature and matter profiles for thin branes and thick branes.

\noindent b)Some of the proposed thick matter sheet IVP's: planar thick branes and spherical thick branes, 
which permit the use of a standard rather than directional notion of asymptotics.

\end{document}